\RequirePackage[table, dvipsnames, svgnames, x11names]{xcolor}

\documentclass[runningheads]{llncs}

\usepackage{eccv}
\usepackage{eccvabbrv}

\usepackage{graphicx}
\usepackage{amsmath}
\usepackage{amssymb}
\usepackage{booktabs}
\usepackage{multirow}
\usepackage{tabu}
\usepackage{arydshln} 
\usepackage{stfloats}
\usepackage{makecell}
\usepackage{listings}
\usepackage{subcaption}
\usepackage{pifont}
\usepackage{adjustbox}
\usepackage{array}
\usepackage{wrapfig}

\usepackage{tcolorbox}
\tcbuselibrary{skins, breakable} 

\definecolor{BrandBlue}{HTML}{1E88E5} 
\definecolor{mygray}{gray}{.9}
\definecolor{brightred}{RGB}{220,0,0}
\definecolor{brightgreen}{RGB}{0,150,0}
\definecolor{cvprblue}{RGB}{0, 115, 189}

\newcommand{\dec}[1]{{\scriptsize\color{brightred}$\downarrow$#1}}

\tcbset{
    promptbox/.style={
        enhanced,
        breakable,
        fonttitle=\bfseries,
        fontupper=\small\rmfamily\linespread{1.1}\selectfont, 
        width=\linewidth,
        top=8pt, bottom=4pt,
        colback=BrandBlue!5!white,
        colframe=BrandBlue!65!black,
        coltext=black,
        colbacktitle=BrandBlue!80!black,
        coltitle=white,
        attach boxed title to top left={yshift=-0.1in,xshift=0.15in},
        boxed title style={boxrule=0pt,colframe=BrandBlue!80!black},
    }
}
\newtcolorbox{PromptBox}[2][]{promptbox, title=#2, #1}

\usepackage[pagebackref,breaklinks,colorlinks,allcolors=black]{hyperref}
\usepackage[accsupp]{axessibility} 
\usepackage{orcidlink}

\begin{document}

\title{ELVA: Exploring Ranking-Driven Universal Multimodal Retrieval} 

\makeatletter
\def\@fnsymbol#1{%
  \ensuremath{%
  \ifcase#1\or \dagger\or \ddagger\or \mathsection\or *\else \@ctrerr\fi}}
\makeatother

\author{Yuhan Liu\inst{1}\thanks{Equal contribution.}\thanks{Work done during the internship at Xiaomi.} \and
Pei Fu\inst{2}\textsuperscript{\ensuremath{\dagger}} \and
Hang Li\inst{2} \and
Yukun Qi\inst{2} \and
Chao Jiang\inst{2} \and
Jingwen Fu\inst{3}\thanks{Corresponding author.}\and
Zhen Liu\inst{1} \and
Bin Qin\inst{2} \and
Zhenbo Luo\inst{2}\and
Jian Luan\inst{2} \and
Jingmin Xin\inst{1}\textsuperscript{\ensuremath{\S}}}

\authorrunning{Y.~Liu et al.}

\institute{National Key Laboratory of Human-Machine Hybrid Augmented Intelligence,\\
Institute of Artificial Intelligence and Robotics, Xi'an Jiaotong University \and
MiLM Plus, Xiaomi Inc \and
Zhongguancun Academy, Beijing, China}

\maketitle


\begin{abstract}
Leveraging Multimodal Large Language Models (MLLMs) via contrastive learning has become a mainstream paradigm for improving the performance of Universal Multimodal Retrieval (UMR).
However, previous works have ignored the \textbf{grain blindness} when adapting the contrastive paradigm into retrieval tasks.
Grain blindness refers to the tendency of the model to overlook grain-level information contained in the query, which is crucial for effectively handling complex queries.
This stems from contrastive learning treating samples as a binary classification (positive/negative), while ignoring the different information carried by each negative sample.
To address this, we argue that negatives should be treated differently according to their similarity to the positive sample, enabling the model to learn distinct grain information from each negative.
In this paper, we introduce a simple but effective framework, called ELVA, a novel rule-based RL framework that mitigates grain blindness through ranking-driven MLLMs. 
1) Instead of relying on reward models, we extend Reinforcement Learning with Verifiable Rewards
 (RLVR) to retrieval tasks, allowing the model to explore new ranking behaviors without explicit ranking labels.
2) By utilizing rule-based rewards, our approach jointly optimizes the ranking of negative samples while enlarging the similarity gap between positive and negative.
To more precisely measure grain blindness, we further introduce MRBench, a new benchmark specifically designed for multi-grain query scenarios. ELVA achieves state-of-the-art results across standard retrieval benchmarks, and its notable 13.1\% improvement on MRBench further demonstrates its effectiveness in alleviating grain blindness.

\end{abstract}    
\section{Introduction}
\label{sec:intro}

Universal Multimodal Retrieval (UMR) refers to a general retrieval paradigm that unifies diverse retrieval tasks within a single framework and enables generalization to unseen retrieval tasks \cite{liu2025lamra,lyu2025puma,zhu2025retrv,kong2025modality}.
This marks a substantial shift from prior efforts, which primarily focused on modality-specific retrieval tasks, including text-to-text \cite{zhao2024dense,ma2024fine}, text-to-image \cite{fu2024linguistic,zhang2020context}, and image-to-image \cite{baldrati2022effective,saito2023pic2word} retrieval.
Recently, researchers have begun exploring Multimodal Large Language Models (MLLMs) \cite{li2024llava,wang2024qwen2,dong2025scalable,xiaomi2025mimo} for UMR, leveraging their extensive pretrained knowledge and strong generalization ability.
Since MLLMs are originally trained for generative objectives (e.g., next-token prediction), recent works have adapted them to retrieval tasks via contrastive learning, effectively transferring their generative abilities to retrieval tasks \cite{gu2025breaking,meng2025vlm2vec,zhang2025qwen3,lin2025sail}.

\begin{figure}[tbp]
    \centering 
    \centerline{\includegraphics[width=1\linewidth]{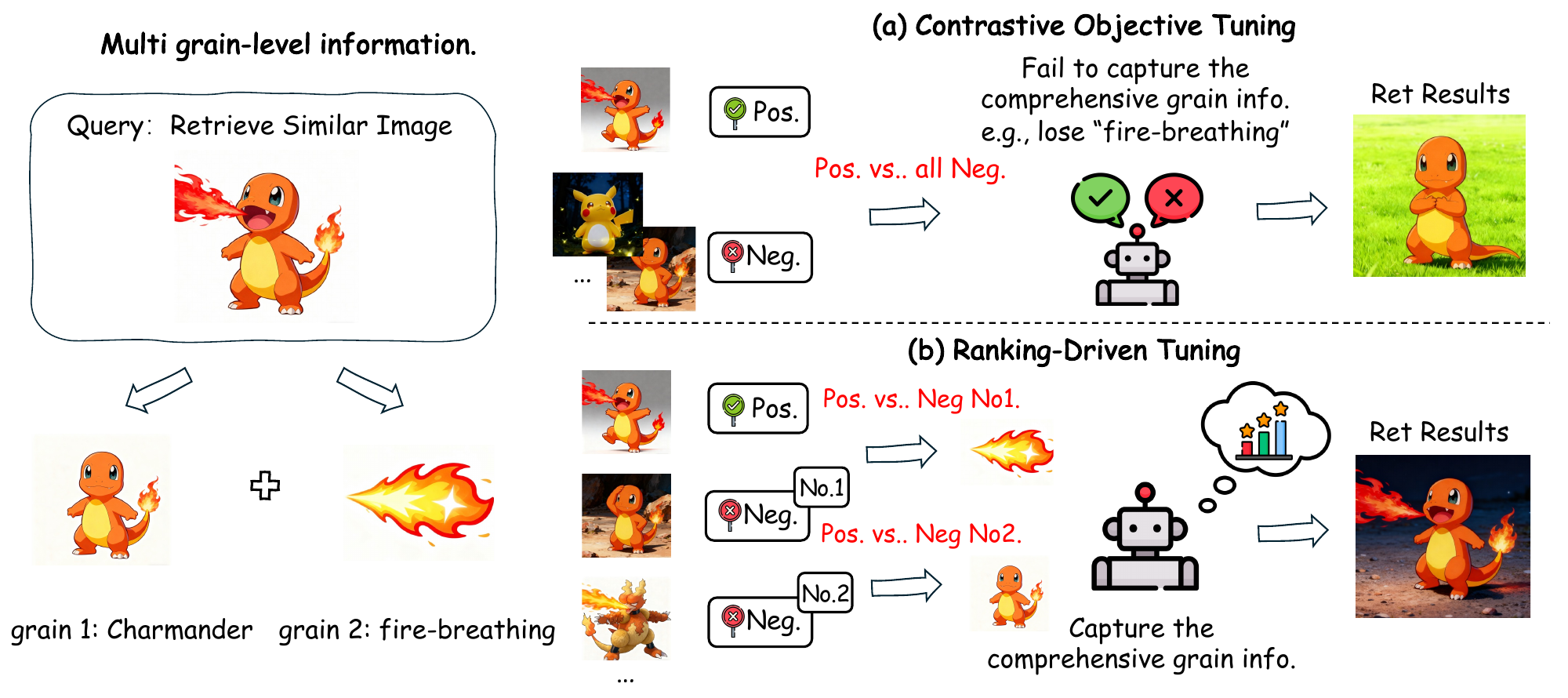}}
    \vspace{-2mm}
    \caption{\textbf{The main idea of our proposed ELVA.} Previous works~\cite{liu2025lamra,lin2024mm} fail on multi-grain queries due to grain blindness that emerges during contrastive training, as illustrated in (a). Build on its basis, our ELVA leverages the ranking-driven tuning with verifiable rewards to capture the comprehensive information, accurately retrieve the precise candidates shown in (b).}
    \label{fig:intro}
    \vspace{-6mm}
\end{figure}

Although existing methods achieve impressive performance, our study reveals that they still suffer from \textit{grain blindness} when adapting the contrastive paradigm to retrieval tasks.
Grain blindness refers to the tendency of the model to overlook grain-level information contained in a query.
Consequently, these methods struggle with multi-grain queries, as shown in Figure \ref{fig:intro}.
In such a case, the query contains multiple levels of grain information, such as the action ``fire-breathing" and the entity ``Charmander", which place high demands on the model’s ability to comprehensively capture multi-grain information.
To tackle the grain blindness, our study highlights two key challenges that need to be confronted.

\textbf{Challenge I:} \quad 
\textit{Which properties that the training paradigm have to equip in order to reduce the grain blindness?}
Previous works \cite{liu2025lamra,lyu2025puma,lin2024mm} leverage contrastive objectives, learning embeddings by distinguishing positive and negative samples. 
However, as illustrated in Figure \ref{fig:intro} (a), this training paradigm is suboptimal for retrieval tasks, as it fails to comprehensively capture the multiple levels of grain information contained in a query.
Intuitively, the positive sample needs to be contrasted against diverse negative samples in order to learn distinct grain information.
Yet, the contrastive paradigm treats all negatives equally \cite{zhu2025generalized,jia2021scaling}, 
ignore the differential information carried by each negative, which is crucial for accurate retrieval.
To address this issue, we argue that the model should treat negative samples differently based on their similarity to the positive, which means negatives with higher similarity should be positioned closer to the positive. 
Toward this goal, we propose a ranking-driven approach that ranks candidates according to their relevance to the positive sample, enabling the model to capture comprehensive grain-level information, as shown in Figure \ref{fig:intro} (b).

\textbf{Challenge II:} \quad 
\textit{How to incentivize the ranking ability without the ranking labels?}
Previous retrieval models typically rely on supervised ranking objectives (e.g., listwise loss) to learn candidate ordering \cite{gu2025unime,qwen3vlembedding,huang2024cosent}. 
However, in UMR scenarios, obtaining ranking labels such as precisely ranking all negative samples by their exact relevance is extremely difficult and expensive. 
Meanwhile, forcing models to fit static labels severely restricts their capacity to explore subtle, grain-level hierarchical differences.
To overcome this, we introduce an exploration-driven RL framework that operates without explicit ranking targets, utilizes the reward function as dynamic evaluators. Driven by the GRPO algorithm \cite{shao2024deepseekmath}, the model autonomously discovers optimal inter-negative hierarchies by comparing the relative ranking quality of its variant generated candidate lists.

In this paper, we propose \textbf{\textit{ELVA}}: \textbf{E}xp\textbf{L}oring Ranking-Driven Uni\textbf{V}ersal Multimodal Retriev\textbf{A}l, a novel rule-based RL framework designed to overcome the grain blindness through ranking-driven MLLMs. 
Different from previous works that rely solely on contrastive objectives \cite{liu2025lamra,lyu2025puma}, 
ELVA adopts a ranking-driven tuning to capture richer grain-level information while simultaneously overcome the absence of ranking labels.
Specifically, we propose two verifiable reward function to optimize the policy model: \textbf{1)} Ranking Reward, which encourages the model to rank candidates based on their relevance while rewarding the model for placing positive samples at higher ranks inspired by the NDCG \cite{jarvelin2002cumulated}. Our Ranking Reward is a \textbf{continuous reformulation for RL}, ensuring continuous reward signal while optimizing negatives hierarchies; \textbf{2)} Margin Reward, which enforces explicit similarity-gap constraints, ensuring that positive samples remain closer to the query than negative ones.
Additionally, we introduce a balanced negative sampling strategy to construct a ranking customized dataset for RL, filtering out excessively difficult negatives to ensure stable optimization.

For a more comprehensive evaluation, we construct a new benchmark, MRBench, derived from the M-BEIR dataset~\cite{wei2023uniir}.
MRBench is specifically designed for multi-grain retrieval, where each query contains two or more grain-level attributes (e.g., an entity and an action), making it particularly challenging to preserve multi-grained information.
Our method achieves a substantial 13.1\% improvement in retrieval accuracy on this benchmark, demonstrating its effectiveness in mitigating grain blindness.

To summarize, we make the following contributions:

\begin{itemize}

\item We identify the issue of grain blindness when adapting the contrastive learning paradigm to retrieval tasks, and highlight two key challenges that need to be confronted.

\item We propose ELVA to enable comprehensive multi-grain information acquisition by jointly optimizing ranking order and enforcing similarity-gap constraints. 

\item We construct a new dataset to evaluate model performance in complex multi-grain scenarios, and ELVA achieves state-of-the-art performance across diverse benchmarks including MRBench.

\end{itemize}

\section{Related Work}
\label{sec:related}

\textbf{Universal Multimodal Retrieval.}
Multimodal retrieval serves as the core task in information retrieval \cite{radford2021learning,jia2021scaling,baldrati2023zero,tang2025missing}, focusing on retrieving related content across diverse data modalities \cite{lee2018stacked,li2021align}.
As the landscape of information retrieval expands, more recent studies have shifted attention toward universal multimodal retrieval (UMR) \cite{liu2025lamra,lyu2025puma,zhu2025retrv,kong2025modality}, where a unified model is capable of handling heterogeneous modalities and diverse retrieval tasks simultaneously.
While earlier work in this domain often relied on small foundation models such as CLIP \cite{wei2024uniir}, recent advances \cite{liu2025lamra,lin2024mm,zhou2025megapairs} have demonstrated the promise of employing Multimodal Large Language Models (MLLMs) \cite{liu2023visual,li2023blip,wang2024qwen2,zhu2023minigpt,qi2026patchcue,yang2026shaping,wu2026dual} to further enhance retrieval performance.
As MLLMs are primarily trained with generative objectives, recent researches \cite{lan2025llave,liu2025lamra,lin2024mm} adapt them for retrieval tasks via contrastive learning, 
utilize embeddings extracted from MLLMs performing similarity-based retrieval, leveraging their strong cross-modal representation capabilities.
However, it remains a significant challenge to capture comprehensive grain-level information in order to retrieve complex queries with high precision. 
We propose a simple yet effective approach to incentivize the model's ranking ability, thus improving the UMR performance.


\vspace{3pt}\noindent \textbf{RL for Ranking Learning.}
Reinforcement learning (RL) has become a promising approach that enables models to adjust their behavior during training based on continuous feedback signals \cite{liu2025visual,yu2024rlhf,shen2025vlm}.
ReasonRank \cite{liu2025reasonrank} optimizes discrete metric-based rewards (e.g., NDCG, Recall, RBO) defined over the entire ranked list. 
MM-R5 \cite{xu2025mm} introduces a position-weighted ranking reward, where each retrieved item receives a score according to its ranking position. 
These discrete rewards which means it only jumps when the ranking order changes, the optimization signal is discontinuous and high-variance, leading to unstable learning and poor convergence \cite{xiao2022towards,zhou2024optimizing}.
In contrast, our reward enforces continuous feedback offering smoother gradients than purely discrete metric-based rewards.
Search-R3 \cite{gui2025search} incorporates continuous similarity scores, however the feedback easily saturates once the top-ranked positions converge, providing weak supervision for representation learning.
In this paper, our ELVA not only maintains non-saturating learning signals via margin reward, but also models the intra-negative ranking structure via continuous ranking reward. Moreover, the ranking reward encourages negatives with high relevance ranked nearer to the positive, thereby enriching the grain-level information within the embedding space.

\section{Method}
\label{sec:method}

\begin{figure*}[ht]
  \centering
  \includegraphics[width=\textwidth]{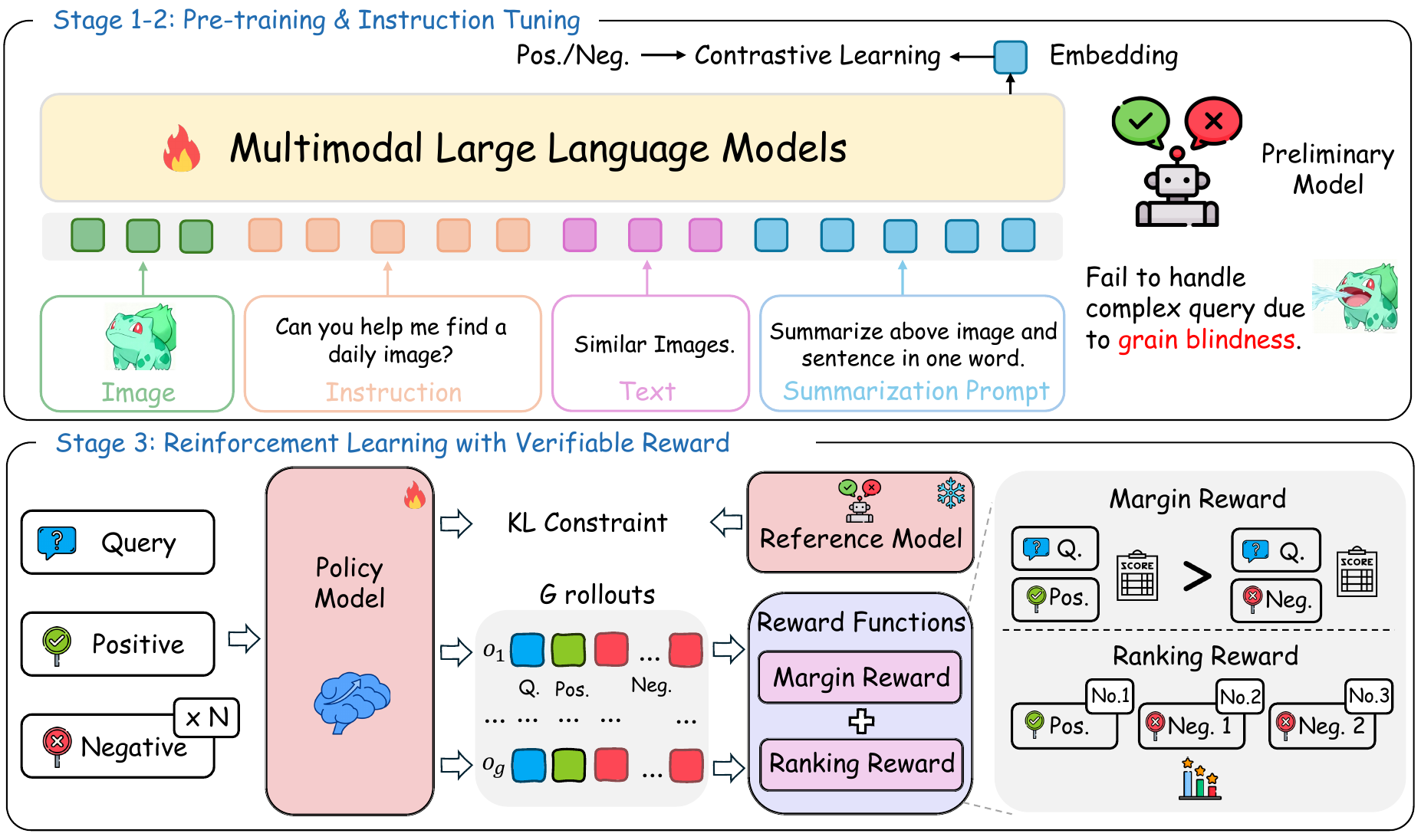} \\
  \vspace{-5pt}
  \caption{\textbf{Overview of the proposed ELVA framework.} ELVA leverage the three-stage training framework for UMR tasks. Stage 1-2 as the pre-training and instruction tuning stage following \cite{liu2025lamra}, obtained the preliminary model struggle with the grain blindness. Stage 3 employ the RL tuning to incentivizing the ranking ability to address the issue. Given the input query $q$ and candidates including pos. and $N$ x neg., we first perform G rollouts to output $G$ independent sets of embeddings
   from policy model. Then we compute the reward $r_{i}$ for each output $o_{i}$ using our proposed reward functions, detailed in Section \ref{sec:3.4}. Finally, we optimize the policy model with GRPO \cite{guo2025deepseek} while ensuring that the model remains close to the reference policy model, via KL divergence.}
  \vspace{-1em}
 \label{fig:arch}
\end{figure*}

\subsection{Preliminary}
\label{sec:3.1}

This paper proposes a novel framework ELVA in Figure \ref{fig:arch}, to tackle the grain blindness in UMR. In this task, given a query $q$ originating from any modality (text, image, or interleaved formats), the objective is to retrieve the most relevant sample from a set $\Omega = \{c_{n}\}_{n=1}^{N}$ with $N$ candidates. To perform retrieval, we use a unified embedding extractor to encode both the query and candidates.
Then we compute the cosine similarity $s_i$ between the query and each candidate, and rank all candidates based on their similarity scores.
The final retrieval output corresponds to the top-$k$ candidates:
\[
\mathcal{C} = \Phi_{\text{ret}}(q, \Omega)
= \operatorname{Top}\text{-}k\big( \{ s_1, s_2, \ldots, s_N \} \big).
\]

\subsection{Formulation}
\label{sec:3.2}
To theoretically ground the concept of grain blindness, we formally define the mathematical formulation and this representational collapse to theoretically ground our framework.

\begin{definition}[Grain and Query Composition]
A grain $g$ is defined as the minimal, atomic semantic unit (e.g., an entity, an attribute, or an action) within a multimodal query. A query $q$ is formulated as a set of these interdependent units, $q=\{g_1, g_2, \dots, g_K\}$. Grains possess inherent structural dependencies; for instance, in the query ``red dress,'' the attribute grain \textit{red} is necessarily coupled with the entity grain \textit{dress} to form a coherent search intent.
\end{definition}

\begin{definition}[Grain Blindness]
\label{d2}
\textbf{Grain blindness} is formalized as a representational collapse where the distance between the full query embedding and its de-grained version (where a specific grain $g_k$ is removed) falls below a discriminative threshold $\delta $:
\begin{equation}
    d(f_\theta(q), f_\theta(q \setminus \{g_k\})) < \delta ,
\end{equation}
where $f_\theta(\cdot)$ denotes the embedding function and $d(\cdot, \cdot)$ is a distance metric. This inequality indicates that the model fails to preserve the discriminative features of the specific grain $g_k$ in the embedding space.
\end{definition}

Theoretically, the emergence of grain blindness during contrastive learning can be attributed to Gradient Starvation~\cite{pezeshki2021gradient,robinson2021can}. If certain dominant grains within $q$ (e.g., a primary entity) provide a sufficient similarity margin to distinguish between positive and negative pairs, the contrastive loss drops rapidly. Consequently, other salient but secondary grains lose their necessary gradient signals for optimization. This premature convergence forces the model to ignore the suppressed grains, directly leading to the representational collapse described in Definition \ref{d2}.

\subsection{Pre-training \& Instruction Tuning}
\label{sec:3.3}


To effectively leverage the contrastive learning paradigm in transforming the generative capability of MLLMs into discriminative representations, we first employ a two-stage training framework following \cite{liu2025lamra,lyu2025puma}.
Since MLLMs are primarily trained for generative objectives such as next-token prediction, their inherent retrieval capability remains limited.
The first stage conducts language-only pretraining on NLI datasets \cite{gao2021simcse}, enabling the generative model to produce more effective embeddings.
The second stage, instruction tuning, further aligns the MLLMs with various retrieval tasks for better adaptability. 
These tasks include image-to-image retrieval, composed image retrieval, and image/question-to-multimodal-document retrieval, among others.
Further details on the instruction templates and the M-BEIR datasets\cite{wei2023uniir} are provided in the Suppl.

\vspace{2pt}
\noindent \textbf{Training Objective.}
We adopt contrastive learning with the InfoNCE loss~\cite{oord2018representation} as the optimization objective for both the language-only pretraining and instruction-tuning phases.
During training, we input query $q$ and instruction $i$ into the model to obtain the query representation $e_q$. Similarly, each candidate $c$ is fed into the model to derive its representation $e_c$. 
The training objective maximizes the similarity of positive samples and minimizes the similarity of negative samples, formulated as:
\begin{equation*}
\mathcal{L}= -\frac{1}{N} \sum_{n=1}^{N}
\log \left[
\frac{\exp \left[ \cos\left(e_{q}, e_{c}^{+}\right) / \tau \right]}
     {\sum_{m=1}^{N} \exp \left[ \cos\left(e_{q}, e_{c_{m}}\right) / \tau \right]}
\right]
\end{equation*}
\noindent where $\tau$ denotes the temperature parameter and $N$ is the batch size. This approach enables the model to learns to discriminate between relevant and irrelevant information across various modalities.

\subsection{ELVA}
\label{sec:3.4}
After the above training stages, we obtain a preliminary retrieval model.
However, the model lacks sufficiently grain-level information, making it less effective for complex queries.
To address this, we propose ELVA, a reinforcement learning based framework incentivizes the ranking ability of MLLMs for UMR.
In contrast to conventional contrastive learning, ELVA optimizes model behavior through rule-based reward signals.
However, a critical challenge in applying Policy Optimization to UMR tasks is that traditional EOL extraction \cite{liu2025lamra,lyu2025puma} (i.e., directly extracting the hidden state of a fixed prompt token \cite{jiang2023scaling}) yields no representational variance, rendering RL exploration impossible. 
To confront this, we formulate the feature extraction process as a \textbf{generative paradigm}. The model is mandated to first autoregressively generate a textual synthesis of the input, followed by a designated special token \texttt{[RET]}, which serves as the information bottleneck. We utilize the following prompt template:

\begin{PromptBox}{Template for Generative Embeddings}
\textbf{USER}: 
\{$I^{image}$\} <Instruction> \{$I^{query}$\}.

Analyze and summarize the key information of the above input. Finally, append the special token \texttt{[RET]} to represent the entire input.

\textbf{ASSISTANT}: 
\{\texttt{Generated Summary}\} \texttt{[RET]}
\end{PromptBox}

Here, $I^{image}$ denotes the image input, while $I^{query}$ refers to the query text input. These modalities can be flexibly combined, and the corresponding instructions are adjusted accordingly.
$\texttt{[RET]}$ is a special token registered in the LLM, and we take the hidden state at this token’s position as the retrieval embedding.
In our RL framework, an action consists of generating embeddings for the query, positive, and all negatives by the policy model shown in the bottom of Figure \ref{fig:arch}. 
To facilitate policy optimization, the model generates $G$ independent rollouts for a given query $q$ and candidate set $\Omega$ through GRPO. This results in $G$ groups of embeddings, denoted as $\{\mathbf{e}_q^{(g)}, \mathbf{e}_{pos}^{(g)}, \{\mathbf{e}_{n,i}^{(g)}\}_{i=1}^N \}_{g=1}^G$, which serve as the basis for computing relative rewards.

\subsubsection{Verifiable Reward Design.}
\label{sec:3.4.1}

The reward model serves as a key component in reinforcement learning (RL), guiding the model’s behavior to align with predefined correctness objectives. While conventional RL paradigms typically depend on human preference-based reward modeling~\cite{liu2024skywork,kaufmann2024survey}, recent advances such as DeepSeek-R1~\cite{guo2025deepseek} have shown that verifiable reward functions can substantially enhance reasoning capability.
Building upon this insight, we extend Reinforcement Learning with Verifiable Rewards (RLVR) to the multimodal retrieval domain by developing a rule-driven, multi-criteria reward function that jointly evaluates ranking quality and distance between candidates.
This design not only produces accurate retrieval results but also encourages the reasonable order among negative samples, thereby improving both robustness and interpretability.
Our framework evaluates model output in terms of two complementary dimensions: \textbf{ranking orders and the distance between positive and negative}.

\vspace{2pt}
\noindent \textbf{Margin Reward.} The margin reward is designed to encourage the model to maintain a sufficient similarity gap between positive and negative samples, inspired by the triplet loss \cite{hong2022unsupervised,liu2024semantic}.
To achieve this, we compute the similarity scores between the query and all negative candidates, and select the hardest negative that has the highest similarity to the query.
The reward then encourages the model to increase the similarity gap between the positive pair and the hardest negative. Given the input of query $q$ and candidates $\Omega = \{c_p, c_{n}^{0} , \cdots, c_{n}^{k}\}$, where p and n represent positive samples and negative samples, the margin reward is defined as:
\begin{equation*}
    R_{\text{Margin}} = \text{max}(0,\text{cos}(q,c_{p})- \text{cos}(q,c_{n}) _{\text{max}}-\delta),
\end{equation*}
where $\delta$ is a predefined hyperparameter that specifies the minimum required similarity gap.
This formulation effectively drives the positive sample to rank at the top of the candidate list while imposing explicit similarity gap constraints. This reward enhances both retrieval precision and the discriminative structure of the learned representations.

\vspace{2pt}
\noindent \textbf{Ranking Reward.}
We propose the ranking reward to explicitly encourage the model to rank the positive sample at the top of the candidate list while simultaneously promoting the ordering among negative samples to capture sufficient grained information. 
In contrast to the margin reward, which enforces a fixed similarity gap between positive and negative pairs, the ranking reward introduces rank-dependent weighting to optimize both rank precision and inter-negative structure.

Given a query and a candidate set, to translate these candidates into a ranked list, we compute cosine similarity scores $s_i = \cos(q, c_i)$ between the query and candidates within each specific rollout. These scores are then sorted to form the ranking used to calculate the reward.
The reward is defined as:
\begin{equation*}
R_{\text{Rank}} = s_{(r)} \cdot \frac{1}{1+\log r} \;-\;
\gamma \sum_{k \neq r} s_{(k)} \cdot (\log k -1),
\end{equation*}
where $s_{(r)}$ denotes the similarity of the positive sample ranked at position $r$, $k$ denotes the rank of negatives and $\gamma$ controls the penalty strength for high-scoring negatives. 
The first term encourages the model to assign a higher similarity score to the positive sample and place it near the top of the ranking. The second term encourages more similar negative samples to be ranked closer to the top, thereby receiving smaller penalties, while still constraining their similarity to the query.

Moreover, $R_{Rank}$ is a continuous reformulation designed for RL, ensuring smoother reward landscape while optimizing negatives hierarchies. 
This rank-aware formulation enhances retrieval quality by encouraging precise rank positioning and structured ranking among negatives, thereby improving the model’s ability to learn adequate grained, discriminative representations.

\vspace{2pt}
\noindent \textbf{Final Reward Function.}
The total reward combines the above rewards to optimize both ranking quality and similarity gap:
\[
R_{\text{total}} = \alpha  R_{\text{Margin}} + \varepsilon  R_{\text{Ranking}},
\]
where hyperparameters $\alpha $ and $\varepsilon $ balance reward contributions and avoid reward hacking. This joint optimization enables the model to produce more precise retrieval outcomes by effectively capturing rich grained information, particularly in the context of complex or compositional queries.

\subsubsection{Negative Sampling Strategy}
We introduce a balanced negative sampling strategy tailored for the RL stage. Prior work~\cite{liu2025lamra} directly sampled the top-100 candidates for SFT, which we find problematic for RL: when candidates are overly similar to the query, the reward distribution becomes too narrow, yielding weak or vanishing gradients~\cite{yu2025dapo} and hindering effective policy learning.
To address this, we construct each query’s candidate set using a balanced mix of negatives:
(1) 50 filtered hard negatives, obtained by removing candidates above a similarity threshold and selecting the top 50 remaining ones~\cite{li2025u,gu2025breaking}; and
(2) 50 randomly sampled negatives from the full pool.
This combination increases reward variance and provides richer learning signals: the filtered subset offers controlled difficulty to enhance ranking capability, while the random subset adds distributional diversity and prevents overfitting.
Overall, this mixed sampling strategy yields a more stable and informative signal for RL optimization.

\label{sec:3.5}

\section{Experiments}
\label{sec:experiments}
\subsection{Experimental Setup}

\vspace{3pt}\noindent \textbf{Datasets and Metrics.}  
We employ the NLI dataset~\cite{gao2021simcse} for pre-training and the M-BEIR dataset~\cite{wei2023uniir} for instruction tuning, following~\cite{liu2025lamra,lyu2025puma}.
M-BEIR spans eight retrieval tasks across ten datasets, containing approximately 1.1M training instances.
For the RL stage, we apply our negative sampling strategy to construct a training set of 11k instances from M-BEIR by sampling 1\% of data from each dataset.
We evaluate ELVA on the M-BEIR test set to assess its versatility across diverse retrieval scenarios.
To further examine its generalization ability, we also evaluate ELVA on several unseen datasets~\cite{zhang2024long,baldrati2023zero}.
For multi-grain scenarios, we introduce a new benchmark, MRBench (Multi-gRain Benchmark), derived from M-BEIR. We first employ Qwen2.5-VL-7B \cite{bai2025qwen2} to automatically identify and filter queries containing at least two grain-level attributes, followed by human sampling verification to ensure data quality. Finally, we sample an equal number of instances from each task, resulting in a benchmark of 1k queries across 3 datasets and 4 retrieval tasks.
We follow standard evaluation protocols for all datasets, using Recall@K as the primary metric for retrieval tasks.

\vspace{2pt}\noindent \textbf{Implementation Details.} 
Our framework is implemented in PyTorch, by default, built upon Qwen2-VL-7B~\cite{wang2024qwen2}.
During the retrieval pretraining stage, experiments are conducted on 8 × H 20 GPUs with a batch size of 576, a learning rate of $4\times10^{-5}$, and trained for two epochs (3h completed).
In the instruction tuning stage, we use 16 × H 20 GPUs with a batch size of 960 and a learning rate of $1\times10^{-4}$ for one epoch following \cite{liu2025lamra} (48h completed).
For the RL stage, training is performed for one epoch on 8 × H 20 GPUs with  a learning rate of $1\times10^{-6}$, using 8 rollouts and $\beta = 0.2$ (16h completed).  
Across all stages, the vision encoder remains frozen, while the language model is fine-tuned using LoRA.
During M-BEIR evaluation, we conduct experiments in a local retrieval pool with generative embedding extract method. 
The weight hyperparameter set to $\alpha=0.4$ and $\varepsilon =0.6$.
More details are shown in Suppl.

\begin{table*}[t]
\caption{\textbf{Comparison with recent state-of-the-art methods on the M-BEIR test set.} The first row denotes the retrieval task configuration, where $q^t$ and $q^i$ represent text and image queries, respectively, and $c^t$ and $c^i$ denote text and image candidates. Dataset abbreviations include VN for VisualNews, F200K for Fashion200K, InfoS for InfoSeek, and FIQ for FashionIQ. Following the UniIR \cite{wei2023uniir} evaluation protocol, Recall@10 is reported for FashionIQ and Fashion200K, while Recall@5 is used for all other datasets. The best results are highlighted.}
\vspace{-7pt}
\centering
\resizebox{\linewidth}{!}{
\begin{tabular}{lc@{\hspace{0.1cm}}c@{\hspace{0.1cm}}c@{\hspace{0.1cm}}c@{\hspace{0.1cm}}c@{\hspace{0.1cm}}c@{\hspace{0.1cm}}c@{\hspace{0.1cm}}c@{\hspace{0.1cm}}c@{\hspace{0.1cm}}c@{\hspace{0.1cm}}c@{\hspace{0.1cm}}c@{\hspace{0.1cm}}c@{\hspace{0.1cm}}c@{\hspace{0.1cm}}c@{\hspace{0.1cm}}c@{\hspace{0.1cm}}c@{\hspace{0.1cm}}}
\toprule
 & \multicolumn{3}{c}{{$q^t \to c^i$}} & {$q^t \to c^t$} & \multicolumn{2}{c}{{$q^t \to (c^i, c^t)$}} & \multicolumn{3}{c}{{$q^i \to c^t$}} & {$q^i \to c^i$} & \multicolumn{2}{c}{{$(q^i, q^t) \to c^t$}} & \multicolumn{2}{c}{{$(q^i, q^t) \to c^i$}} & \multicolumn{2}{c}{{$(q^i, q^t) \to (c^i, c^t)$}} & \\
 \cmidrule(r){2-4} \cmidrule(r){5-5}  \cmidrule(r){6-7} \cmidrule(r){8-10} \cmidrule(r){11-11} \cmidrule(r){12-13} \cmidrule(r){14-15} \cmidrule(r){16-17} 
 \textbf{Methods} & VN  & COCO & F200K & WebQA & EDIS & WebQA & VN & COCO & F200K & NIGHTS & OVEN & InfoS & FIQ & CIRR & OVEN & InfoS & \textbf{Avg.} \\
\cmidrule(r){2-4} \cmidrule(r){5-5}  \cmidrule(r){6-7} \cmidrule(r){8-10} \cmidrule(r){11-11} \cmidrule(r){12-13} \cmidrule(r){14-15} \cmidrule(r){16-17} 
& R@5 & R@5 & R@10 & R@5 & R@5 & R@5 & R@5 & R@5 & R@10 & R@5 & R@5 & R@5 & R@10 & R@5 & R@5 & R@5 & \\
\midrule
\rowcolor{gray!12}
\multicolumn{18}{c}{\textit{\textbf{Zero-shot}}} \\
\midrule
CLIP-L~\cite{radford2021learning} & 43.3 & 61.1 & 6.6 & 36.2 & 43.3 & 45.1 & 41.3 & 79.0  & 7.7 & 26.1 & 24.2 & 20.5 & 7.0 & 13.2 & 38.8 & 26.4 & 32.5    \\
SigLIP~\cite{zhai2023sigmoid} & 30.1 & 75.7 & 36.5 & 39.8 & 27.0 & 43.5 & 30.8 & 88.2  & 34.2 & 28.9 & 29.7 & 25.1 & 14.4 & 22.7 & 41.7 & 27.4 & 37.2  \\
BLIP~\cite{li2022blip} & 16.4 & 74.4 & 15.9 & 44.9 & 26.8 & 20.3 & 17.2 & 83.2  & 19.9 & 27.4 & 16.1 & 10.2 & 2.3 & 10.6 & 27.4 & 16.6 & 26.8  \\
BLIP2~\cite{li2023blip} & 16.7 & 63.8 & 14.0 & 38.6 & 26.9 & 24.5 & 15.0 & 80.0  & 14.2 & 25.4 & 12.2 & 5.5 & 4.4 & 11.8 & 27.3 & 15.8 & 24.8  \\
Qwen2-VL-7B~\cite{wang2024qwen2} & 9.3 & 55.1 & 5.0 & 42.0 & 26.2 & 9.4 & 5.4 & 46.6 & 4.0 & 21.3 & 21.4 & 22.5 & 4.3 & 16.3 & 43.6 & 36.2 & 23.0 \\
Qwen2.5-VL-7B~\cite{bai2025qwen2} & 40.2 & 71.9 & 20.3 & 71.9 & 49.4 & 64.5 & 29.3 & 84.6 & 19.4 & 25.5 & 42.4 & 32.1 & 25.0 & 55.1 & 60.8 & 54.9 & 46.7 \\
\midrule
\rowcolor{gray!12}
\multicolumn{18}{c}{\textit{\textbf{Supervised - Dual Encoder}}} \\
\midrule
$\text{UniIR-BLIP}_{\text{FF}}$~\cite{wei2024uniir} & 23.4 & 79.7 & 26.1 & 80.0 & 50.9 & 79.8 & 22.8 & 89.9 & 28.9 & 33.0 & 41.0 & 22.4 & 29.2 & 52.2 & 55.8 & 33.0 & 46.8  \\
$\text{UniIR-CLIP}_{\text{SF}}$~\cite{wei2024uniir} & 42.6 & 81.1 & 18.0 & 84.7 & 59.4 & 78.7 & 43.1 & 92.3 & 18.3 & 32.0 & 45.5 & 27.9 & 24.4 & 44.6 & 67.6 & 48.9 & 50.6  \\
\midrule
\rowcolor{gray!12}
\multicolumn{18}{c}{\textit{\textbf{Supervised - MLLMs}}} \\
\midrule

Vision-R1-7B~\cite{huang2025vision}& 41.9 & 75.0 & 22.0 & 70.6 & 51.3 & 69.1 & 35.4 & 85.1 & 22.4 & 25.9 & 48.8 & 44.0 & 29.2 & 57.7 & 66.2 & 59.0 & 50.2 \\
VLM-R1-7B~\cite{shen2025vlm}& 40.5 & 77.2 & 22.5 & 72.3 & 50.0 & 67.9 & 36.2 & 86.3 & 20.9 & 26.4 & 48.8 & 37.5 & 29.9 & 57.4 & 64.0 & 62.3 & 50.0 \\
MM-Embed-7B~\cite{lin2024mm}& 41.0 & 71.3 & 17.1 & \textbf{95.9} & \textbf{68.8} & 85.0 & 41.3 & 90.1 & 18.4 & 32.4 & 42.1 & 42.3 & 25.7 & 50.0 & 64.1 & 57.7 & 52.7 \\
PUMA-3B~\cite{lyu2025puma}& 35.7 & 79.5 & 25.8 & 86.2 & 58.2 & 78.4 & 35.2 & 90.1 & 29.0 & 31.4 & 52.7 & 48.3 & 30.6 & 49.9 & 74.0 & 65.2 & 54.4 \\
LamRA-Ret-2B \cite{liu2025lamra} & 30.8 & 78.8 & 23.1 & 82.5 & 54.3 & 77.8 & 31.2 & 88.5 & 27.1 & 28.7 & 51.1 & 44.2 & 28.9 & 47.7 & 72.3 & 60.8 & 51.6 \\
LamRA-Ret-7B \cite{liu2025lamra} & 41.6 & 81.5 & 28.7 & 86.0 & 62.6 & 81.2 & 39.6 & 90.6 & 30.4 & 32.1 & 54.1 & 52.1 & 33.2 & 53.1 & 76.2 & 63.3 & {56.6} \\
\midrule
ELVA-2B (Ours) & 35.6 & 80.3 & 25.0 & 88.0 & 56.1 & 80.5 & 33.4 & 90.2 & 25.9 & 29.3 & 52.0 & 47.4 & 30.9 & 50.0 & 72.8 & 61.3 & 53.8$_{\textcolor{SeaGreen}{+4.3\%}}$ \\

ELVA-7B (Ours) & \textbf{43.5} & \textbf{83.0} & \textbf{29.2} & 91.0 & 63.5 & \textbf{83.1} & \textbf{41.7} & \textbf{92.2} & \textbf{32.1} & \textbf{32.8} & \textbf{56.0} & \textbf{55.5} & \textbf{34.6} & \textbf{55.4} & \textbf{77.5} & \textbf{67.1} & \textbf{58.7}$_{\textcolor{SeaGreen}{+3.9\%}}$ \\

\bottomrule
 \end{tabular}
}
\vspace{-10pt} 
\label{tab:mbeir}
\end{table*}

\subsection{Experimental Results}

\noindent \textbf{Comparison of Effectiveness.}
We begin by evaluating the effectiveness of ELVA on the M-BEIR test set.
Table~\ref{tab:mbeir} reports results in terms of Recall@K, covering 16 sub-tasks across 8 combinations of query and candidate modalities.
To examine scalability, we present results for both ELVA-2B and ELVA-7B.
We compare against three categories of methods:
1) Zero-shot general-purpose MLLMs, including BLIP-2~\cite{li2023blip} and Qwen-VL~\cite{wang2024qwen2};
2) prior RL-based MLLMs, such as Vision-R1~\cite{huang2025vision} and VLM-R1~\cite{shen2025vlm}; and
3) Retrieval-specialized MLLMs, including MM-Embed~\cite{lin2024mm} and LamRA~\cite{liu2025lamra}.
As shown in Table~\ref{tab:mbeir}, ELVA consistently achieves state-of-the-art (SOTA) results across most settings.
Notably,
1) even the 2B variant surpasses larger models such as MM-Embed-7B on most sub-tasks,
and 2) on particularly challenging configurations like $(q^{i}, q^{t}) \rightarrow (c^{i},c^{t})$ on the InfoS dataset with 6.0\% improvement, ELVA attains a substantial performance lead over previous methods.
These results demonstrate the robustness and universality of ELVA, highlighting its strong retrieval capability across diverse multimodal inputs.
We further apply LamRA-Rank in the reranking stage to boost accuracy shown in Suppl.

\begin{table*}[t]
\caption{\textbf{Experimental results on unseen datasets.} The first row denotes the type of retrieval task: $q^t$ represents text queries, $q^i$ image queries, $q^{\text{dialog}}$ dialog-based queries, and $(q^i \oplus q^t)$ denotes interleaved image-text queries; $c^t$ and $c^i$ correspond to text and image candidates, respectively, while ITM refers to the Image-Text Matching task. Dataset abbreviations include Share4V for ShareGPT4V, Urban for Urban-1k, VisD for Visual Dialog, and MT-FIQ for Multi-round FashionIQ.
The $^*$ symbol indicates that images in these datasets originate from COCO or FashionIQ; however, due to notable differences in captions and query structures, they are still treated as unseen datasets. We follow the standard evaluation metrics defined for each dataset, and the best-performing results are highlighted.}
\vspace{-6pt}
\setlength{\tabcolsep}{1.5mm}
\centering
\resizebox{\linewidth}{!}{
\begin{tabular}{lcccccccccccc}
\toprule
 & \multicolumn{3}{c}{{$q^t \to c^i$}} & \multicolumn{3}{c}{{$q^i \to c^t$}} & \multicolumn{2}{c}{{$(q^i, q^t) \to c^i$}} & \multicolumn{1}{c}{{$q^{\text{dialog}} \to c^i$}} & 
\multicolumn{1}{c}{{$(q^i \oplus q^t) \to c^i$}} &\multicolumn{2}{c}{{ITM}}\\
\cmidrule(r){2-4} \cmidrule(r){5-7}  \cmidrule(r){8-9} \cmidrule(r){10-10}  \cmidrule(r){11-11} \cmidrule(r){12-13} 
Methods & Share4V  & Urban$^*$ & Flickr & Share4V  & Urban$^*$ & Flickr  & CIRCO$^*$ & GeneCIS$^*$ & VisD$^*$ & MT-FIQ$^*$ & CC-Neg & Sugar-Crepe$^*$ \\
\cmidrule(r){2-4} \cmidrule(r){5-7}  \cmidrule(r){8-9} \cmidrule(r){10-10}  \cmidrule(r){11-11} \cmidrule(r){12-13} 
& R@1 & R@1 & R@1 & R@1 & R@1 & R@1 & MAP@5 & R@1 & R@1 & R@5 & Acc. & Acc. \\
\midrule
CLIP-L~\cite{radford2021learning} & 84.0 & 52.8 & 67.3 & 81.8 & 68.7 & 87.2 & 4.0 & 13.3 & 23.7 & 17.7 & 66.7 & 73.0 \\
Long-CLIP-L~\cite{zhang2024long} & {95.6} & 86.1 & 76.1 & \textbf{{95.8}} & 82.7 & 89.3 & 5.7 & 16.3 & 37.9 & 18.5 & 76.3 & 80.9 \\
UniIR-CLIP~\cite{wei2023uniir} & 85.8 & 75.0 & 78.7 & 84.1 & 78.4 & 94.2 & 12.5 & 16.8 & 26.8 & 39.4 & 79.9 & 80.3 \\
E5-V~\cite{jiang2024e5} & 86.7 & 84.0 & 79.5 & 84.0 & 82.4 & 88.2 & 24.8 & 18.5 & 54.6 & 19.2 & {83.2} & 84.7 \\
MagicLens-L~\cite{zhang2024magiclens} & 85.5 & 59.3 & 72.5 & 60.9 & 24.2 & 84.6 & 29.6 & 16.3 & 28.0 & 22.6 & 62.7 & 75.9 \\
EVA-CLIP-8B~\cite{sun2024eva} & 91.2 & 77.8 & 80.8 & 93.1 & 80.4 & 95.6 & 6.0 & 13.1 & 23.2 & 22.1 & 59.4 & 81.7\\
EVA-CLIP-18B~\cite{sun2024eva} & 92.1 & 81.7 & {83.3} & 94.0 & 83.3 & \textbf{{96.7}} & 6.1 & 13.6 & 24.7 & 21.9 & 63.8 & 83.1\\
LamRA-Ret-7B \cite{liu2025lamra} & 93.3 & 95.1 & 82.8 & 88.1 & 94.3 & 92.7 & 33.2 & 18.9 & 62.8 & 60.9 & 79.6 & 85.8 \\
\midrule
\rowcolor{gray!15}
ELVA-7B (Ours) &\textbf{ 96.6} & \textbf{96.1} & \textbf{84.4} & 92.0 & \textbf{95.5} & 95.2 & \textbf{34.5} & \textbf{20.2} & \textbf{65.3} & \textbf{61.2} & \textbf{87.3} & \textbf{91.1} \\
\bottomrule
 \end{tabular}
}
\label{tab:zero-shot}
\vspace{-10pt}
\end{table*}

\begin{table}[t]
  \begin{minipage}[c]{0.56\linewidth}
    \centering
    \caption{\footnotesize \textbf{Experimental results on held-out tasks.} $^*$ indicates training on other tasks without exposure to the three held-out tasks.}
    \setlength{\tabcolsep}{1.1mm}
    \resizebox{1\linewidth}{!}{
      \begin{tabular}{lcccccc}
        \toprule
         & {$q^i \to c^i$} & \multicolumn{2}{c}{{$(q^i, q^t) \to c^t$}} & \multicolumn{2}{c}{{$(q^i, q^t) \to (c^i, c^t)$}}\\
         \cmidrule(r){2-2} \cmidrule(r){3-4}  \cmidrule(r){5-6}  
         Methods & NIGHTS & OVEN & InfoS & OVEN & InfoS & Avg. \\
        & R@5 & R@5 & R@5 & R@5 & R@5 & \\
        \midrule
        \multicolumn{7}{l}{\color{gray}{\textit{Supervised}}} \\
        \addlinespace[0.15em]
        $\text{UniIR-BLIP}_{\text{FF}}$ & \textbf{33.0} & 41.0 & 22.4 & 55.8 & 33.0 & 37.0 \\
        $\text{UniIR-CLIP}_{\text{SF}}$ & 32.0 & 45.5 & 27.9 & \textbf{67.6} & 48.9 & 44.4 \\
        \midrule
        \multicolumn{7}{l}{\color{gray}{\textit{Zero-shot}}} \\
        \addlinespace[0.15em]
        Qwen2.5-VL & 20.3 & 38.5 & 40.4 & 53.6 & 44.9 & 39.5 \\   
        Vision-R1 & 22.9 & 39.8 & 42.9 & 57.4 & 46.5 & 41.9 \\
        LamRA-Ret$^*$ & 27.2 & 44.7 & 44.0 & 62.8 & 49.5 & 45.6 \\
        \midrule
        \rowcolor{gray!15}
        ELVA-7B$^*$ & 28.2 & \textbf{46.5} &\textbf{49.2} & 64.4 & \textbf{53.0} & \textbf{48.3} \\
        \bottomrule
      \end{tabular}
    }
    \label{tab:task_generalization}
  \end{minipage}
  \hfill
  \begin{minipage}[c]{0.41\linewidth}
    \centering
    \caption{\footnotesize \textbf{Comparison of Reward Weighting.}}
    \vspace{-4pt}
    \resizebox{1\linewidth}{!}{
      \begin{tabular}{lcccc}
        \toprule
        \textbf{Method} & \textbf{VN} & \textbf{COCO} & \textbf{F200K} & \textbf{Avg.}\\
        \midrule
        $\alpha=0.6$, $\varepsilon =0.4$ & 42.9  & 82.2 & 28.8  & 58.2\\
        $\alpha=0.5$, $\varepsilon =0.5$ & 43.2  & 82.5 & 29.1  & 58.4\\
        \rowcolor{gray!15}
        $\alpha=0.4$, $\varepsilon =0.6$ (ELVA) & \textbf{43.5}  & \textbf{83.0} & \textbf{29.2}  & \textbf{58.7}\\
        \bottomrule
      \end{tabular}
    }
    \label{tab:weight}
    
    \vspace{5pt} 
    
    \centering
    \caption{\footnotesize \textbf{Generalizability of our ranking-driven RL framework.}}
    \vspace{-6pt}
    \resizebox{1\linewidth}{!}{
      \begin{tabular}{lcc}
        \toprule
         \textbf{Method} & \textbf{M-BEIR}  & \textbf{MRBench} \\
        \midrule
         PUMA\cite{lyu2025puma} & 54.4 & 35.1  \\
         \rowcolor{gray!15}
         PUMA+ELVA  & \textbf{56.3} & \textbf{37.0} \\
        
         MM-Embed\cite{lin2024mm} & 52.7  & 34.6 \\
         \rowcolor{gray!15}
         MM-Embed+ELVA & \textbf{54.9} & \textbf{36.2} \\
        \bottomrule
      \end{tabular}
    }
    \label{tab:gen}
  \end{minipage}
  \vspace{-8pt}
\end{table}

\vspace{1pt}
\noindent \textbf{Comparison of Generalization on Unseen Dataset.}
To assess the generalization capability of our approach, we conduct extensive experiments on multiple \textit{\textbf{unseen retrieval datasets}}.
As shown in Table~\ref{tab:zero-shot}, 
ELVA consistently delivers strong performance across all evaluation settings, demonstrating robust generalization to diverse data modalities and task types.
In ITM tasks, our ELVA achieves over a 9.7\% improvement to other methods. 
Similarly, in fixed-modal retrieval tasks such as text-to-image, our method also achieves substantial improvements in performance.
These findings underscore the strong adaptability and scalability of ELVA, highlighting its promise as a unified framework for broader multimodal applications.

\vspace{1pt}
\noindent \textbf{Comparison of Generalization on Unseen Task.}
Evaluate ELVA on \textbf{\textit{unseen retrieval tasks}} by excluding specific tasks during training and testing the retrained model on these omitted tasks. As shown in Table~\ref{tab:task_generalization}, our method exhibits strong performance on unseen retrieval tasks. At the same parameter scale, our ELVA achieves 5.9\% improvement over previous SOTA method.
This strong generalization capability indicates that our method can effectively extend to unseen retrieval tasks without further training, showing great potential for broader real-world applications.

\begin{figure*}[t]
  \centering
  \includegraphics[width=\textwidth]{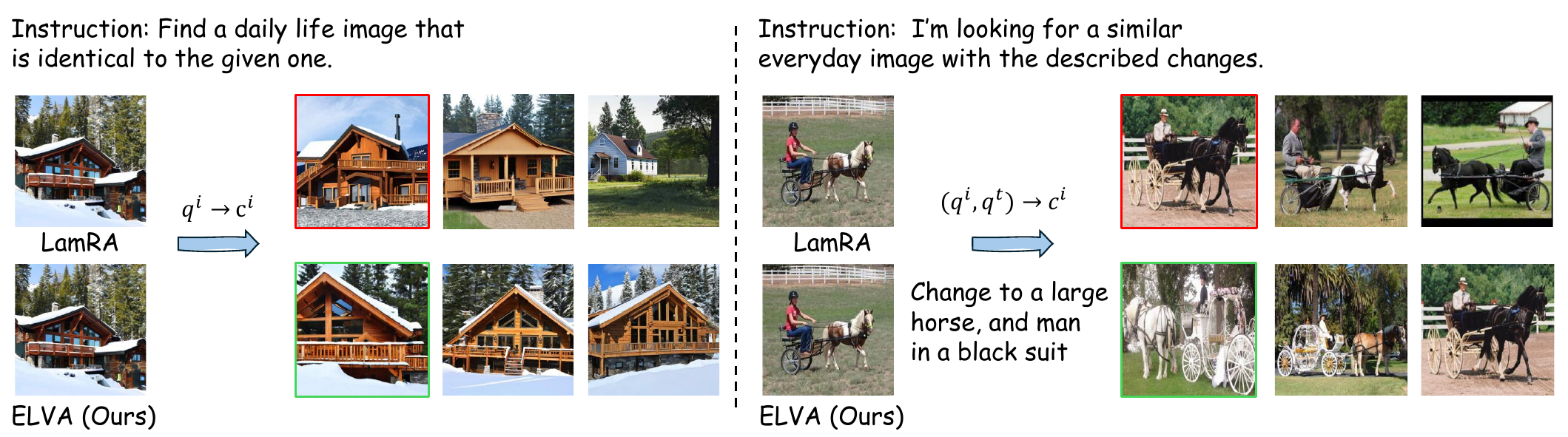} \\
  \vspace{-4pt}
  \caption{\textbf{Qualitative examples.} We show the results of our method across different retrieval tasks, with the correct result indicated by the green box. Here, $q^t$ for text queries, $q^i$ for image queries, $c^i$ for image candidates.}
    \vspace{-13pt}
 \label{fig:qualitative}
\end{figure*}

\vspace{1pt}
\noindent \textbf{Comparison of Multi-Grain scene on MRBench.}
Table \ref{tab:MRBench} reports the results on the MRBench benchmark.
ELVA achieves superior performance compared to previous methods SOTA LamRA with 13.1\% improvements and zero-shot models, highlighting the effectiveness of our approach in accurately retrieving queries with multi-grain information. 
As illustrated in Figure \ref{fig:qualitative}, 
when a query includes multiple grain-level attributes, such as ``snow-capped mountains’’ and ``trees’’, existing methods frequently fail to retrieve the correct result due to their inability to capture all grain components adequately.
The qualitative comparisons show that our method captures the complex intent of the query and successfully retrieves the desired target. 
These results further confirm the effectiveness of our proposed approach in substantially alleviating the grain blindness problem.
For more qualitative examples, please refer to Suppl.

\begin{table}[t]
\begin{minipage}{0.48\linewidth}
\centering
\caption{\footnotesize \textbf{Experimental results on MRBench datasets.} The $^*$ symbol indicates that dataset are filtered for the multi-grain scene.} 
\vspace{-2pt}
\setlength{\tabcolsep}{1.5mm}
\centering
\resizebox{1\linewidth}{!}{
\begin{tabular}{lccccc}
\toprule
   & {$q^t \to c^i$} & {$q^i \to c^t$} & {{$q^i \to c^i$}} & {{$(q^i, q^t) \to c^i$}} & \\
 \cmidrule(r){2-2} \cmidrule(r){3-3} \cmidrule(r){4-4}  \cmidrule(r){5-5}  
 Methods & COCO$^*$ & COCO$^*$ & NIGHTS$^*$  & CIRR$^*$ & Avg. \\
& R@5 & R@5 & R@5  & R@5 & \\
\midrule
\addlinespace[0.15em]
Qwen2-VL-7B \cite{wang2024qwen2}  & 39.2 & 32.0 & 15.6  & 10.8 & 34.3  \\
LamRA-Ret-7B \cite{liu2025lamra}& 50.4 & 54.0 & 22.4  & 25.7 & 38.1  \\
\midrule
\rowcolor{gray!15}
ELVA-7B (Ours) & \textbf{55.5} & \textbf{60.6} &\textbf{24.1}  & \textbf{32.5} & \textbf{43.2} \\
\bottomrule
 \end{tabular}
}
\label{tab:MRBench}
\end{minipage}%
\hfill
\begin{minipage}{0.48\linewidth}
\centering
\caption{\footnotesize \textbf{Zero-shot text-to-video retrieval performance.}}
\setlength\tabcolsep{1.5mm}
\centering
\resizebox{1\linewidth}{!}{
    \begin{tabular}{lcccccc}
        \toprule
        \multirow{2}{*}{\bf{Method}} & \multicolumn{3}{c}{\bf{MSR-VTT}} & \multicolumn{3}{c}{\bf{MSVD}} \\
        \cmidrule(r){2-4} \cmidrule(r){5-7}
         &  R@1 & R@5 & R@10 & R@1 & R@5 & R@10  \\
         \midrule
        InternVideo~\cite{wang2022internvideo} & 40.0 & 65.3 & 74.1 & 43.4 & 69.9 & 79.1 \\
        ViCLIP~\cite{wang2023internvid} & 42.4 & - & - & 49.1 & - & - \\
        UMT-L~\cite{li2023unmasked} & 42.6 & 64.4 & 73.1 & 49.9 & 77.7 & 85.3 \\
        InternVideo2$_{s2}$-6B~\cite{wang2024internvideo2} & 55.9 & 78.3 & 85.1 & 59.3 & 84.4 & 89.6\\
    LamRA-7B \cite{liu2025lamra} & 44.7 & 68.6 & 78.6 & 52.4 & 79.8 & 87.0\\
    \midrule
    \rowcolor{gray!15}
    ELVA-7B (Ours) & \textbf{46.4} & \textbf{70.5} & \textbf{78.9} & \textbf{53.9} & \textbf{80.7} & \textbf{87.9}\\
      \bottomrule
      \end{tabular}
}
\label{tab:zero_shot_video_ret}
\end{minipage}
\vspace{-8pt}
\end{table}

\vspace{-15pt}
\begin{table}[t]
\caption{
    \textbf{Ablation study.} 
    Avg. refers to the average recall performance across the M-BEIR test set. We select the image-to-text retrieval tasks as example.
    }
    \vspace{-6pt}
    \centering
    \renewcommand\arraystretch{1}
    \setlength{\tabcolsep}{2.0mm}

    \resizebox{0.9\linewidth}{!}{
\begin{tabular}{clcccc}

    \toprule
    \textbf{\#} &
	\multirow{1}[0]{*}{\textbf{Method}} & 

	\multicolumn{1}{c}{ \textbf{VN} } &
	\multicolumn{1}{c}{ \textbf{COCO}} &
	\multicolumn{1}{c}{\textbf{F200K}} &
	\multicolumn{1}{c}{\textbf{Avg.}} 
	 \\

    \midrule
	1 & w/o Ranking Rewards           & 41.7 \dec{1.8} & 82.0 \dec{1.0} & 28.4 \dec{0.8} & 57.2 \dec{1.5} \\
    2 & w/o Margin Rewards              & 42.3 \dec{1.2}& 82.2 \dec{0.8} & 28.5 \dec{0.7} & 58.1 \dec{0.6} \\
    3 & w/o Negative Ranking            & 42.8 \dec{0.7}& 82.0 \dec{1.0} & 28.5 \dec{0.7} & 58.1 \dec{0.6} \\
	4 & w/o Negative Sampling Strategy  & 43.1 \dec{0.4}& 82.2 \dec{0.8}  & 28.5 \dec{0.7}  & 58.2 \dec{0.5}  \\
    5 & w/o Randomly Sampled Negatives  & 43.3 \dec{0.2} & 82.6 \dec{0.4} & 28.9 \dec{0.3}  & 58.4 \dec{0.3} \\
    \midrule
    \rowcolor{gray!15}
    6 &ELVA-7B (Ours)  & \textbf{43.5} &  \textbf{83.0}  & \textbf{29.2} & \textbf{58.7}  \\
    \bottomrule
\end{tabular}
}
    \label{tab:abla_reward_func}
    \vspace{-10pt}
\end{table}

\subsection{Ablation Study}
\noindent \textbf{Ablating Reward Functions.}
To evaluate the effect of reward functions, we ablate ranking rewards and margin rewards, analyzing their effect across M-BEIR test set, as shown in Table \ref{tab:abla_reward_func}. 
Excluding the ranking rewards leads to noticeable performance drops, highlighting its importance for enhancing the ranking quality, in order to learn sufficient grain information.
Removing the margin rewards also reduces performance, indicating its role in ensuring the similarity gap for precise retrieval.
In addition, we conduct another ablation in ranking reward, removing the second term designed for negative rank in row 3.
The performance indicates that optimizing the negative ranking leads to more accurate retrieval. 
We conduct more ablation studies about hyperparameters in Suppl.

\vspace{4pt}
\noindent \textbf{Ablating Negative Sampling Strategy.}
To evaluate the impact of the negative sampling strategy, we analyze different sampling strategies. 
As shown in Table \ref{tab:abla_reward_func}, removing the sampling strategy leads to performance degradation, suggesting its importance in maintaining training data moderate difficulty in row 4.
Introducing a random subset further enhances distributional diversity and prevents the model from overfitting in row 5. 
The results indicate that a properly training dataset is essential for model convergence, ensuring robust performance across datasets.
More comparison results are shown in Suppl.

\vspace{2pt}
\noindent \textbf{Ablating Reward Weighting.}
Additionally, we examine how different weighting schemes between the margin-based reward and the ranking-based reward affect the final performance in Table \ref{tab:weight}. This observation indicates that while the margin reward provides effective local pairwise guidance, it is the ranking reward that captures more holistic ordering signals and aligns more strongly with the final evaluation metric (Recall/K). Therefore, assigning a slightly higher weight to the ranking reward allows the model to better optimize global ranking behavior without losing the corrective constraints introduced by the margin term.

\vspace{-5mm}
\subsection{Deep Analysis}
\vspace{-2mm}
\noindent\textbf{Post-Training Extension.} 
Beyond the complete three stages pipeline, our proposed RL paradigm (Stage 3) functions as a highly adaptable, modular enhancement for existing multimodal retrievers. Future works can bypass the supervised fine-tuning stages and directly apply our ranking-driven RL to off-the-shelf models to achieve consistent performance gains. As shown in Table \ref{tab:gen} integrating S3 as a post-training step significantly improves the average performance of PUMA \cite{lyu2025puma} and MM-Embed\cite{lin2024mm}. This demonstrates our RL framework to be a universal "plug-and-play" booster that seamlessly scales to various multimodal retrieval architectures.

\vspace{1pt}
\noindent \textbf{Extending to Video Retrieval.}
As presented in Table~\ref{tab:zero_shot_video_ret}, we evaluate our method on the MSR-VTT~\cite{xu2016msr} and MSVD~\cite{chen2011collecting} datasets under a zero-shot text-to-video retrieval setting.
The results show that our approach achieves strong performance.
For example, on MSR-VTT, our model achieves 16.5\% improvements over InternVideo~\cite{wang2022internvideo} , while on MSVD, it outperforms UMT-L~\cite{li2023unmasked} by 8.4\%.
It is noteworthy that our model has not been exposed to any video data during fine-tuning, yet it still retains Qwen2-VL’s inherent video understanding ability.
Although the current performance remains below the state-of-the-art InternVideo2~\cite{wang2024internvideo2}, we plan to incorporate video data in future work to further narrow this gap \cite{qi2025vcr,Wu_2025_CVPR}.

\begin{wrapfigure}{r}{0.55\textwidth}
  \vspace{-24pt} 
  \centering
  \includegraphics[width=\linewidth]{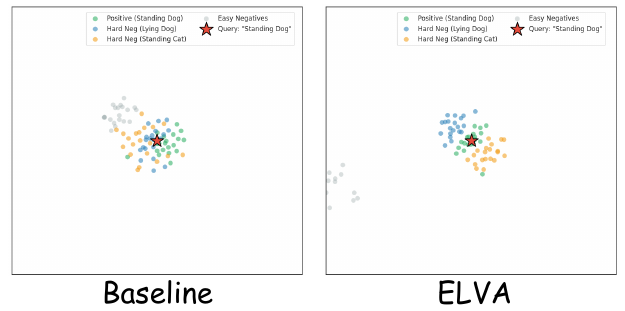}
  \vspace{-7mm}
  \caption{Distribution of embeddings from  Baseline (left) and ELVA (right).}
  \label{fig:tsne_visualization}
  \vspace{-20pt} 
\end{wrapfigure}
\vspace{2pt}
\noindent \textbf{Empirical and Qualitative Analysis of the Embedding Space.}
To validate the mitigation of grain blindness (Def. \ref{d2}), we measure the representation distance $d(f_{\theta}(q), f_{\theta}(q \setminus \{g_k\}))$ on 100 sampled MRBench queries. By systematically masking a single phrase-level grain (e.g., dropping `standing'' from `standing dog''), we compute the average cosine distance between full and masked query embeddings. The baseline yields a collapsed distance of 0.07, indicating the dropped grain was largely ignored. Conversely, ELVA significantly widens this gap to 0.15, quantitatively confirming its preservation of grain-level semantics. Furthermore, t-SNE visualization in Figure \ref{fig:tsne_visualization} of the query `Standing Dog'', shows that while the baseline entangles positives with hard negatives (e.g., `Lying Dog'' or ``Standing Cat''), ELVA clearly separates them. Together, these results demonstrate ELVA's ability to prevent granularity loss and construct a highly discriminative embedding space.

\section{Limitations and Future Work}

The limitation of MLLM-based retrieval is high inference costs. This overhead can be further mitigated via feature precomputation, layer pruning \cite{lyu2025puma}, efficient MLLM designs \cite{zhang2025llava,wu2024cr,wu2026hierarchical}, or deploying the lightweight ELVA-2B. Future works includes exploring joint retrieval-reranking frameworks to reduce pipeline complexity and scaling to larger MLLMs.

\section{Conclusion}
In this paper, we present ELVA, a novel framework designed to address the grain blindness when adapting the MLLMs via contrastive paradigm to Universal Multimodal Retrieval (UMR). 
We identify two challenges for addressing the grain blindness: 1) training paradigm's properties impact on retrieval; 2) how to incentivize the new ability without the ranking labels. 
We also introduce a new benchmark specifically designed to evaluate the grain blindness mitigation.
Extensive experiments demonstrate that ELVA achieves significant gains and effectively mitigates grain blindness. 
We expect that our framework will offer meaningful guidance for advancing multimodal information retrieval and encourage continued exploration in this domain.

\section*{Acknowledgement}
This work is supported by the Zhongguancun Academy (Grant No. XTS0070).



%
%
\bibliographystyle{splncs04}
\bibliography{main}

\appendix
\clearpage

\section{Review of GRPO}
The GRPO algorithm originally introduced in DeepSeekMath~\cite{guo2025deepseek}, is a reinforcement learning framework designed to enhance reasoning capabilities of large language models (LLMs) without relying on a separate critic network—a major limitation in traditional algorithms such as Proximal Policy Optimization (PPO)~\cite{schulman2017proximal}. Conventional approaches like PPO employ a value function to estimate the expected quality of generated responses, which often introduces instability and substantial computational overhead. In contrast, GRPO removes the dependency on a value estimator by directly performing relative comparisons among groups of responses, offering a more efficient and stable alternative for large-scale language model fine-tuning\cite{huang2026exposurebiasalleviatedirectional}.

Given a question $q$, the old policy model $\pi_{\theta_{\text{old}}}$ generates a group of $G$ candidate responses ${o_1, o_2, \dots, o_G}$. Each response $o_i$ is then evaluated by a rule-based reward function $R(q, o_i)$, yielding a corresponding scalar reward $r_i$, defined as:
\begin{equation}
    r_i = R(q, o_i) = 
    \begin{cases} 
    1 ,& \text{if } o_i \text{ = ground truth}, \\
    0 ,& \text{otherwise},
    \end{cases} 
\end{equation}
where $R(\cdot, \cdot)$ is a verifiable reward function that compares the model output against the ground truth under task-specific rules. In our work, we extend this formulation by designing multi-objective reward functions specialized for retrieval tasks, which encourage the model to optimize ranking orders while maintaining the similarity gap between the positive sample and negatives. 

Instead of relying on absolute reward values, GRPO adopts a relative normalization strategy within each response group. The advantage for the $i$-th response is computed as:
\begin{align}
A_i = \frac{r_i - \text{mean}({r_1, \dots, r_G})}{\text{std}({r_1, \dots, r_G})},
\end{align}
where $A_i$ reflects the relative quality of $o_i$ compared to other responses in the same group. The sequence-level advantage $A_i$ is then uniformly applied to all decoding steps of the corresponding output sequence. This normalization eliminates the need for a critic model while ensuring stable and efficient updates.
The objective of GRPO is to maximize the relative advantages of generated responses while constraining the updated policy $\pi_{\theta}$ to remain close to a reference policy $\pi_{\text{ref}}$. Accordingly, the GRPO loss is formulated as:
\begin{align}
\mathcal{L}_{\text{GRPO}}(\theta) &= -\frac{1}{G}\sum_{i=1}^G \frac{1}{\left|o_i\right|} \sum_{t=1}^{\left|o_i\right|}  \bigg[  
\frac{\pi_{\theta}(o_{i,t} \mid q, o_{i,\textless t})}{ \varphi\big[\pi_{\theta}(o_{i,t} \mid q, o_{i,\textless t} ) \big] } A_{i,t}  \notag \\
&\quad - \beta \mathbb{D}_{\text{KL}}(\pi_{\theta} \parallel \pi_{\text{ref}}) \bigg] ,
\label{eq:simply_grpo}
\end{align}
where $\varphi[\cdot]$ denotes the stop-gradient operation, $\beta$ controls the strength of the KL regularization, and $\mathbb{D}_{\text{KL}}(\pi{\theta} \parallel \pi_{\text{ref}})$ penalizes deviations from the reference policy to prevent catastrophic forgetting. This design allows GRPO to efficiently align LLM behavior with task-specific objectives while maintaining stability and sample efficiency during fine-tuning.

\section{More Implementation Details}

\vspace{2pt}

\vspace{2pt}
\subsection{Supplement on RL Training Dataset.}
For the RL stage, we uniformly sample 1\% of data from each dataset and apply our negative sampling strategy to construct an 11k-instance training set from M-BEIR.

\subsection{Details of MRBench Construction}
To systematically identify multi-grain queries for the construction of MRBench, we employ Qwen2.5-VL-7B \cite{bai2025qwen2} as an automatic multimodal parser. Given that the queries in our benchmark encompass diverse modalities including pure text, pure images, and interleaved image-text formats. We leverage the strong multimodal comprehension capabilities of Qwen2.5-VL-7B to seamlessly process these inputs. The model is tasked with decomposing each candidate query into atomic semantic units (grains) following Definition 1. To ensure strict programmatic filtering, we constrain the model to output a standardized JSON format.

\vspace{0.5em}
\noindent\textbf{Prompt Design.} We design a zero-shot prompt that explicitly instructs the model to extract and categorize key structural grains into entities and actions from both visual and textual modalities. Notably, we deliberately exclude simple attributes (e.g., colors or sizes) to ensure the benchmark focuses on true compositional complexity rather than trivial descriptive variations. The exact prompt template is provided below:

\begin{PromptBox}{Prompt for Multimodal Grain Extraction}
\textbf{System:} You are an expert multimodal parser specializing in retrieval queries. Your task is to decompose the given query (which may contain images, text, or both) into fine-grained structural semantic units (grains). You must output the result strictly in valid JSON format, without any additional explanations.

\textbf{User:} Please extract the core structural semantic grains from the following multimodal query. Categorize them into two distinct lists: "entities" (objects, persons, or locations) and "actions" (verbs, motions, or dynamic activities). Do NOT include simple attributes like colors, sizes, or static states. Extract information from both the visual and textual content if present.

Query:
[Image Input] (if applicable)
[Text Input] (if applicable)

Output Format:
\{
  "entities": ["..."],
  "actions": ["..."]
\}
\end{PromptBox}

\vspace{0.5em}
\noindent\textbf{Filtering Criteria \& Thresholds.} After generating the JSON responses, we programmatically parse the output to evaluate the structural granularity of each query. A query is classified as a "multi-grain" query and admitted to the initial MRBench candidate pool only if it satisfies the following strict threshold:
\begin{equation}
    |entities| \ge 1 \quad \text{and} \quad |entities| + |actions| \ge 2
\end{equation}
By explicitly excluding attributes from the count, this criterion ensures that MRBench filters out trivial queries (e.g., "a red car") and strictly isolates queries with high compositional complexity. A passing query must contain either multiple interacting objects (e.g., "a man with a dog", yielding 2 entities) or a specific entity engaged in a dynamic behavior (e.g., "a dog running", yielding 1 entity and 1 action). 

\vspace{0.5em}
\noindent\textbf{Quality Assurance.} To verify the reliability of our automated filtering process and ensure high data quality, we incorporated a human verification step. We randomly sampled 100 queries from the filtered subset for manual inspection. The human evaluation confirmed that all sampled queries correctly met the strict structural multi-grain criteria without any issues, demonstrating the accuracy and robustness of our automated pipeline. The final 1k MRBench dataset was then uniformly sampled from this verified pool across different datasets and tasks.

\begin{table*}[h]
\caption{\textbf{Comparison with up-to-date state-of-the-arts on M-BEIR test set in global pool setting.} The first row indicates the retrieval task type: $q^t$ for text queries, $q^i$ for image queries, $c^t$ for text candidates, and $c^i$ for image candidates. Abbreviations used include VN for VisualNews, F200K for Fashion200K, InfoS for InfoSeek, and FIQ for FashionIQ. Evaluation standards follow UniIR, with FashionIQ and Fashion200K using Recall@10, while all other evaluations employ Recall@5.} 
\centering
\resizebox{\linewidth}{!}{
\begin{tabular}{lc@{\hspace{0.1cm}}c@{\hspace{0.1cm}}c@{\hspace{0.1cm}}c@{\hspace{0.1cm}}c@{\hspace{0.1cm}}c@{\hspace{0.1cm}}c@{\hspace{0.1cm}}c@{\hspace{0.1cm}}c@{\hspace{0.1cm}}c@{\hspace{0.1cm}}c@{\hspace{0.1cm}}c@{\hspace{0.1cm}}c@{\hspace{0.1cm}}c@{\hspace{0.1cm}}c@{\hspace{0.1cm}}c@{\hspace{0.1cm}}c@{\hspace{0.1cm}}}
\toprule
 & \multicolumn{3}{c}{{$q^t \to c^i$}} & {$q^t \to c^t$} & \multicolumn{2}{c}{{$q^t \to (c^i, c^t)$}} & \multicolumn{3}{c}{{$q^i \to c^t$}} & {$q^i \to c^i$} & \multicolumn{2}{c}{{$(q^i, q^t) \to c^t$}} & \multicolumn{2}{c}{{$(q^i, q^t) \to c^i$}} & \multicolumn{2}{c}{{$(q^i, q^t) \to (c^i, c^t)$}} & \\
 \cmidrule(r){2-4} \cmidrule(r){5-5}  \cmidrule(r){6-7} \cmidrule(r){8-10} \cmidrule(r){11-11} \cmidrule(r){12-13} \cmidrule(r){14-15} \cmidrule(r){16-17} 
 Methods & VN  & COCO & F200K & WebQA & EDIS & WebQA & VN & COCO & F200K & NIGHTS & OVEN & InfoS & FIQ & CIRR & OVEN & InfoS & Avg. \\
\cmidrule(r){2-4} \cmidrule(r){5-5}  \cmidrule(r){6-7} \cmidrule(r){8-10} \cmidrule(r){11-11} \cmidrule(r){12-13} \cmidrule(r){14-15} \cmidrule(r){16-17} 
& R@5 & R@5 & R@10 & R@5 & R@5 & R@5 & R@5 & R@5 & R@10 & R@5 & R@5 & R@5 & R@10 & R@5 & R@5 & R@5 & \\
\midrule
\rowcolor{gray!15}
\multicolumn{18}{c}{\textit{Supervised - Dual Encoder}} \\
\midrule
$\text{UniIR-BLIP}_{\text{FF}}$~\cite{wei2023uniir} & 23.0 & 75.6 & 25.4 & 79.5 & 50.3 & 79.7 & 21.1 & 88.8 & 27.6 & 33.0 & 38.7 & 19.7 & 28.5 & 51.4 & 57.8 & 27.7 & 45.5  \\
$\text{UniIR-CLIP}_{\text{SF}}$~\cite{wei2023uniir} & 42.6 & 77.9 & 17.8 & 84.7 & 59.4 & 78.8 & 42.8 & 92.3 & 17.9 & 32.0 & 39.2 & 24.0 & 24.3 & 43.9 & 60.2 & 44.6 & 48.9  \\
\midrule
\rowcolor{gray!15}
\multicolumn{18}{c}{\textit{Supervised - MLLMs}} \\
\midrule
PUMA-3B \cite{liu2025lamra} & 35.1 & 73.5 & 25.5 & 85.6 & 34.8 & 88.2 & 27.8 & 30.8 & 58.0 & 77.8 & 47.7 & 45.4 & 30.0 & 46.1 & 70.1 & 60.3 & 52.3 \\
LamRA-Ret-7B \cite{liu2025lamra} & 41.3 & 75.4 & 28.7 & 85.8 & \textbf{62.5} & 81.0 & 39.3 & 90.4 & 30.4 & 32.1 & 48.4 & 48.7 & 33.1 & 50.5 & 70.0 & 60.0 & 54.9 \\
ELVA-7B (Ours)  & \textbf{42.0} & \textbf{76.4} & \textbf{28.9} & \textbf{90.8} & 61.8 & \textbf{82.4} & \textbf{39.6} & \textbf{91.4} & \textbf{31.0} & \textbf{32.6} & \textbf{48.5} & \textbf{51.3} & \textbf{33.3} & \textbf{51.4} & \textbf{70.2} & \textbf{63.1} & \textbf{55.8} \\

\bottomrule
 \end{tabular}
}
\label{tab:mbeir_global}
\end{table*}

\begin{table*}[h]
\caption{\textbf{Comparison with up-to-date state-of-the-arts on M-BEIR test set with second-stage reranking setting.}} 
\centering
\resizebox{\linewidth}{!}{
\begin{tabular}{lc@{\hspace{0.1cm}}c@{\hspace{0.1cm}}c@{\hspace{0.1cm}}c@{\hspace{0.1cm}}c@{\hspace{0.1cm}}c@{\hspace{0.1cm}}c@{\hspace{0.1cm}}c@{\hspace{0.1cm}}c@{\hspace{0.1cm}}c@{\hspace{0.1cm}}c@{\hspace{0.1cm}}c@{\hspace{0.1cm}}c@{\hspace{0.1cm}}c@{\hspace{0.1cm}}c@{\hspace{0.1cm}}c@{\hspace{0.1cm}}c@{\hspace{0.1cm}}}
\toprule
 & \multicolumn{3}{c}{{$q^t \to c^i$}} & {$q^t \to c^t$} & \multicolumn{2}{c}{{$q^t \to (c^i, c^t)$}} & \multicolumn{3}{c}{{$q^i \to c^t$}} & {$q^i \to c^i$} & \multicolumn{2}{c}{{$(q^i, q^t) \to c^t$}} & \multicolumn{2}{c}{{$(q^i, q^t) \to c^i$}} & \multicolumn{2}{c}{{$(q^i, q^t) \to (c^i, c^t)$}} & \\
 \cmidrule(r){2-4} \cmidrule(r){5-5}  \cmidrule(r){6-7} \cmidrule(r){8-10} \cmidrule(r){11-11} \cmidrule(r){12-13} \cmidrule(r){14-15} \cmidrule(r){16-17} 
 Methods & VN  & COCO & F200K & WebQA & EDIS & WebQA & VN & COCO & F200K & NIGHTS & OVEN & InfoS & FIQ & CIRR & OVEN & InfoS & Avg. \\
\cmidrule(r){2-4} \cmidrule(r){5-5}  \cmidrule(r){6-7} \cmidrule(r){8-10} \cmidrule(r){11-11} \cmidrule(r){12-13} \cmidrule(r){14-15} \cmidrule(r){16-17} 
& R@5 & R@5 & R@10 & R@5 & R@5 & R@5 & R@5 & R@5 & R@10 & R@5 & R@5 & R@5 & R@10 & R@5 & R@5 & R@5 & \\
\midrule
\rowcolor{gray!15}
\multicolumn{18}{c}{\textit{Supervised - Retrieval}} \\
\midrule
LamRA-Ret-7B \cite{liu2025lamra} & 41.6 & 81.5 & 28.7 & 86.0 & 62.6 & 81.2 & 39.6 & 90.6 & 30.4 & 32.1 & 54.1 & 52.1 & 33.2 & 53.1 & 76.2 & 63.3 & {56.6} \\
ELVA-7B (Ours) & \textbf{43.5} & \textbf{83.0} & \textbf{29.2} & \textbf{91.0} & \textbf{63.5} & \textbf{83.1} & \textbf{41.7} & \textbf{92.2} & \textbf{32.1} & \textbf{32.8} & \textbf{56.0} & \textbf{55.5} & \textbf{34.6} & \textbf{55.4} & \textbf{77.5} & \textbf{67.1} & \textbf{58.7} \\
\midrule
\rowcolor{gray!15}
\multicolumn{18}{c}{\textit{Supervised - Retrieval + Reranking}} \\
\midrule
LamRA-Ret-7B + LamRA-Rank \cite{liu2025lamra} & 48.0 & \textbf{85.2} & 32.9 & 96.7 & \textbf{75.8} & 87.7 & 48.6 & 92.3 & \textbf{36.1} & 33.5 & 59.2 & 64.1 & \textbf{37.8} & 63.3 & 79.2 & 78.3 & 63.7 \\
ELVA-7B (Ours) + LamRA-Rank\cite{liu2025lamra}  & \textbf{49.0} & \textbf{85.2} & \textbf{33.5} & \textbf{97.4} & 75.6 & \textbf{88.0} & \textbf{49.5} & \textbf{92.5} & 36.0 & \textbf{34.3} & \textbf{59.5} & \textbf{64.3} & 37.6 & \textbf{63.9} & \textbf{79.8} & \textbf{78.5} & \textbf{64.0} \\

\bottomrule
 \end{tabular}
}
\label{tab:mbeir_rank}
\end{table*}

\section{Additional Experimental Results}
\subsection{Experimental Results on the M-BEIR in Global Pool Setting} 

The M-BEIR benchmark supports two evaluation configurations: the global pool and the local pool. The primary distinction lies in how the candidate pool is constructed—either aggregating candidates from all datasets or limiting them to those belonging to the dataset currently being evaluated.
In the main paper, we present results under the \textbf{local pool configuration}, while this section reports additional findings under the \textbf{global pool configuration}.
Table~\ref{tab:mbeir_global} summarizes the global-pool results. Our approach continues to deliver strong performance in this more challenging setting, achieving a significant improvement over LamRA-Ret \cite{liu2025lamra,du2026pansharpening,du2026fame}.

\subsection{Experimental Results on the M-BEIR with Reranking}
Following the setup in LamRA \cite{liu2025lamra}, we employ the second-stage LamRA-Rank module to further refine our retrieval performance shown in Table \ref{tab:mbeir_rank}. The additional performance gains further validate the effectiveness of our approach, demonstrating that our framework can serve as a plug-and-play module that seamlessly integrates with existing retrieval systems \cite{lin2024mm,lyu2025puma,liu2026essence}.

However, we observe that the performance improvement in the reranking stage is noticeably smaller than that achieved in the retrieval stage. We attribute this to the fact that reranking is designed to reorder candidate items to further enhance retrieval precision, whereas our ELVA has already internalized ranking capabilities during training. This implicitly replaces part of the reranker’s functionality, causing the overall gains to be constrained by the reranker’s limited capacity. Therefore, we suggest that future work could explore augmenting the reranking model with additional abilities, such as reasoning ability to unlock further improvements in retrieval performance \cite{zhu2025retrv,lan2025ume}.

\subsection{Experimental Results with Qwen2-VL Vs. Qwen2.5-VL}

In Table \ref{tab:mbeir_qwen25} and Table \ref{tab:unseen_qwen25}, we conduct a comparative evaluation of Qwen2-VL and Qwen2.5-VL, revealing that our framework consistently delivers strong results on both, which highlights its broad adaptability and robust generalizability\cite{liu2026instruction,liu2026structured,du2026unsupervised}.

\begin{table*}[ht]
\caption{\textbf{Performance comparison between Qwen2-VL and Qwen2.5-VL on the M-BEIR test set.}} 
\centering
\resizebox{\linewidth}{!}{
\begin{tabular}{lc@{\hspace{0.1cm}}c@{\hspace{0.1cm}}c@{\hspace{0.1cm}}c@{\hspace{0.1cm}}c@{\hspace{0.1cm}}c@{\hspace{0.1cm}}c@{\hspace{0.1cm}}c@{\hspace{0.1cm}}c@{\hspace{0.1cm}}c@{\hspace{0.1cm}}c@{\hspace{0.1cm}}c@{\hspace{0.1cm}}c@{\hspace{0.1cm}}c@{\hspace{0.1cm}}c@{\hspace{0.1cm}}c@{\hspace{0.1cm}}c@{\hspace{0.1cm}}}
\toprule
 & \multicolumn{3}{c}{{$q^t \to c^i$}} & {$q^t \to c^t$} & \multicolumn{2}{c}{{$q^t \to (c^i, c^t)$}} & \multicolumn{3}{c}{{$q^i \to c^t$}} & {$q^i \to c^i$} & \multicolumn{2}{c}{{$(q^i, q^t) \to c^t$}} & \multicolumn{2}{c}{{$(q^i, q^t) \to c^i$}} & \multicolumn{2}{c}{{$(q^i, q^t) \to (c^i, c^t)$}} & \\
 \cmidrule(r){2-4} \cmidrule(r){5-5}  \cmidrule(r){6-7} \cmidrule(r){8-10} \cmidrule(r){11-11} \cmidrule(r){12-13} \cmidrule(r){14-15} \cmidrule(r){16-17} 
 Methods & VN  & COCO & F200K & WebQA & EDIS & WebQA & VN & COCO & F200K & NIGHTS & OVEN & InfoS & FIQ & CIRR & OVEN & InfoS & Avg. \\
\cmidrule(r){2-4} \cmidrule(r){5-5}  \cmidrule(r){6-7} \cmidrule(r){8-10} \cmidrule(r){11-11} \cmidrule(r){12-13} \cmidrule(r){14-15} \cmidrule(r){16-17} 
& R@5 & R@5 & R@10 & R@5 & R@5 & R@5 & R@5 & R@5 & R@10 & R@5 & R@5 & R@5 & R@10 & R@5 & R@5 & R@5 & \\
\midrule
\rowcolor{gray!15}
\multicolumn{18}{c}{\textit{Qwen2-VL-7B}} \\
\midrule
LamRA-Ret-7B \cite{liu2025lamra} & 41.6 & 81.5 & 28.7 & 86.0 & 62.6 & 81.2 & 39.6 & 90.6 & 30.4 & 32.1 & 54.1 & 52.1 & 33.2 & 53.1 & 76.2 & 63.3 & {56.6} \\
ELVA-7B (Ours) & \textbf{43.5} & \textbf{83.0} & \textbf{29.2} & \textbf{91.0} & \textbf{63.5} & \textbf{83.1} & \textbf{41.7} & \textbf{92.2} & \textbf{32.1} & \textbf{32.8} & \textbf{56.0} & \textbf{55.5} & \textbf{34.6} & \textbf{55.4} & \textbf{77.5} & \textbf{67.1} & \textbf{58.7} \\
\midrule
\rowcolor{gray!15}
\multicolumn{18}{c}{\textit{Qwen2.5-VL-7B}} \\
\midrule
LamRA-Ret-7B \cite{liu2025lamra} & 38.5 & 81.1 & 24.0 & 85.8 & 59.5 & 81.1 & 36.3 &  91.2 &  23.2 &  30.9 & 58.4 &  57.3 & 32.0 & 52.3 & 79.7 & 65.1 & 56.0 \\
ELVA-7B (Ours)

& \textbf{40.3} & \textbf{82.7} & \textbf{24.6} & \textbf{90.0} & \textbf{61.5} & \textbf{82.4} & \textbf{39.8} & \textbf{91.7} & \textbf{34.6} & \textbf{32.0} & \textbf{60.3} & \textbf{59.9} & \textbf{33.0} & \textbf{55.4} & \textbf{82.0} & \textbf{68.6} & \textbf{58.0} \\

\bottomrule
 \end{tabular}
}
\label{tab:mbeir_qwen25}
\end{table*}

\begin{table*}[ht]
\caption{\textbf{Performance comparison between Qwen2-VL and Qwen2.5-VL on unseen dataset.}}
\centering
\resizebox{\linewidth}{!}{
\begin{tabular}{lc@{\hspace{0.1cm}}c@{\hspace{0.1cm}}c@{\hspace{0.1cm}}c@{\hspace{0.1cm}}c@{\hspace{0.1cm}}c@{\hspace{0.1cm}}c@{\hspace{0.1cm}}c@{\hspace{0.1cm}}c@{\hspace{0.1cm}}c@{\hspace{0.1cm}}c@{\hspace{0.1cm}}c@{}}
\toprule
 & \multicolumn{3}{c}{{$q^t \to c^i$}} & \multicolumn{3}{c}{{$q^i \to c^t$}} & \multicolumn{2}{c}{{$(q^i, q^t) \to c^i$}} & \multicolumn{1}{c}{{$q^{\text{dialog}} \to c^i$}} & 
\multicolumn{1}{c}{{$(q^i \oplus q^t) \to c^i$}} &\multicolumn{2}{c}{{ITM}}\\
\cmidrule(r){2-4} \cmidrule(r){5-7}  \cmidrule(r){8-9} \cmidrule(r){10-10}  \cmidrule(r){11-11} \cmidrule(r){12-13} 
Methods & Share4V  & Urban$^*$ & Flickr & Share4V  & Urban$^*$ & Flickr  & CIRCO$^*$ & GeneCIS$^*$ & VisD$^*$ & MT-FIQ$^*$ & CC-Neg & Sugar-Crepe$^*$ \\
\cmidrule(r){2-4} \cmidrule(r){5-7}  \cmidrule(r){8-9} \cmidrule(r){10-10}  \cmidrule(r){11-11} \cmidrule(r){12-13} 
& R@1 & R@1 & R@1 & R@1 & R@1 & R@1 & MAP@5 & R@1 & R@1 & R@5 & Acc. & Acc. \\
\midrule
\rowcolor{gray!15}
\multicolumn{13}{c}{\textit{Qwen2-VL-7B}}\\
\midrule
LamRA-Ret-7B\cite{liu2025lamra} & 93.3 & 95.1 & 82.8 & 88.1 & 94.3 & 92.7 & 33.2 & 18.9 & 62.8 & 60.9 & 79.6 & 85.8 \\
ELVA-7B (Ours) &\textbf{ 96.6} & \textbf{96.1} & \textbf{84.4} & 92.0 & \textbf{95.5} & \textbf{95.2} & \textbf{34.5} & \textbf{20.2} & \textbf{65.3} & \textbf{61.2} & \textbf{87.3} & \textbf{91.1} \\
\midrule
\rowcolor{gray!15}
\multicolumn{13}{c}{\textit{Qwen2.5-VL-7B}}\\
\midrule
LamRA-Ret-7B\cite{liu2025lamra}  & 93.8 & 95.6 & 82.7 & 92.9 & 96.0 & 93.3 & 34.4 & 19.2 &  64.1 & 60.7 & 78.0 & 86.4 \\
ELVA-7B (Ours) & \textbf{96.8} & \textbf{96.5} & \textbf{84.3} & \textbf{95.8} & \textbf{97.2} & \textbf{95.8} & \textbf{35.0} & \textbf{20.5} & \textbf{66.6} & \textbf{60.9} & \textbf{85.3} & \textbf{92.2} \\
\bottomrule
\end{tabular}
}
\label{tab:unseen_qwen25}
\end{table*}

\section{Details about M-BEIR Datasets}
\subsection{M-BEIR Dataset Composition}
To facilitate a clearer understanding of the UMR setting, we provide additional details of the M-BEIR dataset in Table \ref{tab:mbeir_dataset}. The table summarizes the constituent datasets associated with each retrieval task. M-BEIR covers eight retrieval task types across ten datasets, containing a total of 5.6 million candidate instances. This table is adapted from UniIR~\cite{wei2024uniir}, readers may refer to the original paper for further information.

\begin{table*}[h]
\centering
\caption{\textbf{Summary of the M-BEIR benchmarks.}}
\scriptsize 
\resizebox{.8\textwidth}{!}{  
\setlength{\tabcolsep}{3mm}{
  \begin{tabular}{lllrrrr}
    \toprule
    \textbf{Task} & \textbf{Dataset} & \textbf{Domain} & \textbf{\# Train} & \textbf{\# Dev} & \textbf{\# Test} & \textbf{\# Pool}\\
    \midrule
    \multirow{3}{*}{$q^t \to c^i$} & VisualNews & News & 99K & 20K & 20K & 542K \\
     & MSCOCO & Misc. & 100K & 24.8K & 24.8K & 5K \\
     & Fashion200K & Fashion & 15K & 1.7K & 1.7K & 201K \\
     \midrule
     \multirow{1}{*}{$q^t \to c^t$} & WebQA & Wiki & 16K & 1.7K & 2.4K & 544K \\
     \midrule
     \multirow{2}{*}{$q^t \to (c^i, c^t)$} & EDIS & News & 26K & 3.2K & 3.2K & 1M \\
     & WebQA & Wiki & 17K & 1.7K & 2.5K & 403K \\
     \midrule
     \multirow{3}{*}{$q^i \to c^t$} & VisualNews & News & 100K & 20K & 20K & 537K \\
     & MSCOCO & Misc. & 113K & 5K & 5K & 25K \\
     & Fashion200K & Fashion & 15K & 4.8K & 4.8K & 61K \\
     \midrule
     \multirow{1}{*}{$q^i \to c^i$} & NIGHTS & Misc. & 16K & 2K & 2K & 40K \\
     \midrule
     \multirow{2}{*}{$(q^i, q^t) \to c^t$} & OVEN & Wiki & 150K & 50K & 50K & 676K \\
     & InfoSeek & Wiki & 141K & 11K & 11K & 611K \\
     \midrule
     \multirow{2}{*}{$(q^i, q^t) \to c^i$} & FashionIQ & Fashion & 16K & 2K & 6K & 74K \\
     & CIRR & Misc. & 26K & 2K & 4K & 21K \\
     \midrule
     \multirow{2}{*}{$(q^i, q^t) \to (c^i, c^t)$} & OVEN & Wiki & 157K & 14.7K & 14.7K & 335K \\
     & InfoSeek & Wiki & 143K & 17.6K & 17.6K & 481K \\
     \midrule
    8 tasks & 10 datasets & 4 domains & 1.1M & 182K & 190K & 5.6M \\
    \bottomrule
  \end{tabular}
}
}
\label{tab:mbeir_dataset}
\end{table*}

\subsection{M-BEIR Dataset Instructions}
As shown in Table \ref{tab:mbeir_instruction}, we provide a subset of the instructions used in UMR, selecting one representative instruction from each dataset for illustration. The M-BEIR benchmark offers four diverse instructions per dataset, designed to capture different phrasings and intents for the same retrieval task. In this section, we display one instruction randomly chosen from the four. During both training and evaluation, one instruction is randomly sampled for each instance and used as the input prompt, encouraging the model to generalize across varying instruction formulations.

\begin{table*}[h]
\caption{\textbf{Summary of the Unseen Dataset.}
}
\centering
\resizebox{.8\textwidth}{!}{  
\setlength{\tabcolsep}{1.5mm}{
  \begin{tabular}{lcccc}
    \toprule
    \textbf{Dataset} & \textbf{Image Source} & \textbf{Task} & \textbf{Query Format} & \textbf{Candidate Format}\\
    \midrule
    \multirow{2}{*}{ShareGPT4V} & \multirow{2}{*}{SA-1B} & $q^t \to c^i $ & \texttt{<long text>} & \texttt{<image>} \\
    &  & $q^i \to c^t $ & \texttt{<image>} & \texttt{<long text>} \\
     \midrule
     \multirow{2}{*}{Urban-1K} & \multirow{2}{*}{MSCOCO} & $q^t \to c^i $ & \texttt{<long text>} & \texttt{<image>} \\
    &  & $q^i \to c^t $ & \texttt{<image>} & \texttt{<long text>} \\
     \midrule
     \multirow{2}{*}{Flickr30K} & \multirow{2}{*}{Flickr} & $q^t \to c^i $ & \texttt{<short text>} & \texttt{<image>} \\
    &  & $q^i \to c^t $ & \texttt{<image>} & \texttt{<short text>} \\
    \midrule 
    CIRCO & MSCOCO unlabeled set & $(q^i, q^t) \to c^i $ & \texttt{<image><relative caption>} & \texttt{<image>} \\
    \midrule
    GeneCIS & MSCOCO & $(q^i, q^t) \to c^i $ & \texttt{<image><relative caption>} & \texttt{<image>} \\
    \midrule
    Visual Dialog & MSCOCO & $q^{\text{dialog}} \to c^i $ & \texttt{<Q$_1$><A$_1$>}$\cdots$\texttt{<Q$_\text{j}$><A$_\text{j}$>} & \texttt{<image>} \\
    \midrule 
    Visual Storytelling & Flickr & $ (q^i \oplus q^t) \to c^i $ & \texttt{<text$_1$><image$_1$>}$\cdots$\texttt{<text$_\text{j}$>} & \texttt{<image>} \\
    \midrule
    \multirow{2}{*}{MT-FIQ} & \multirow{2}{*}{FashionIQ} & \multirow{2}{*}{$ (q^i \oplus q^t) \to c^i $} & \multirow{2}{*}{\parbox{7cm}{\texttt{<image$_1$><relative caption$_1$>}$\cdots\\$\texttt{<image$_\text{j}$><relative caption$_\text{j}$>}}} & \multirow{2}{*}{\texttt{<image>}} \\
    & & & & \\
    \midrule
    CC-Neg & CC3M & ITM & \texttt{<image>} & \texttt{<text>} 
    \\
    \midrule
    Sugar-Crepe & MSCOCO & ITM & \texttt{<image>} & \texttt{<text>} \\
    \bottomrule
  \end{tabular}
}
}
\label{tab:unseen_dataset}
\end{table*}

\section{Details about Unseen Dataset}
Here, we provide a detailed overview of the unseen benchmarks in Table~\ref{tab:unseen_dataset}. Although several of these datasets originate from MSCOCO or FashionIQ, their caption styles, annotation protocols, or query compositions differ substantially from the originals.
Hence, we still categorize them as unseen datasets for evaluation.
For example, Urban1K features long-form captions produced by GPT-4V \cite{openai2023gpt4v}, while CIRCO adopts a composed-query format that pairs a reference image with a relative textual modification.
Such differences introduce notable discrepancies from the source datasets, resulting in a meaningful distribution shift.

\section{More Qualitative Results}
\begin{figure*}[t]
  \centering
  \includegraphics[width=\textwidth]{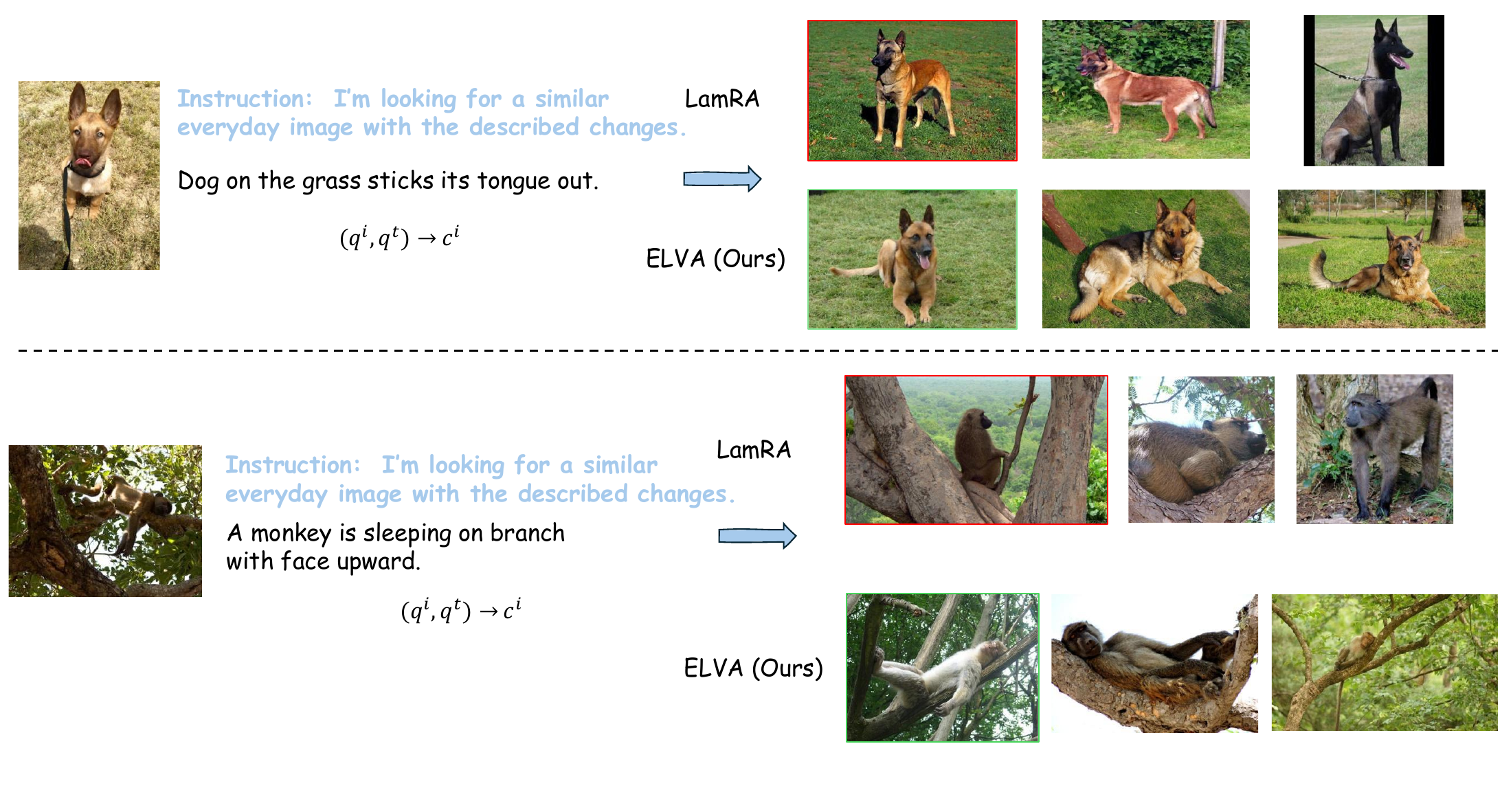} \\
  \vspace{-4pt}
  \caption{\textbf{Qualitative examples.} We show the results of our method across different retrieval tasks, with the correct result indicated by the green box. Here, $q^t$ for text queries, $q^i$ for image queries, $c^i$ for image candidates.}
 \label{fig:suppl1}
\end{figure*}

\begin{figure*}[t]
  \centering
  \includegraphics[width=\textwidth]{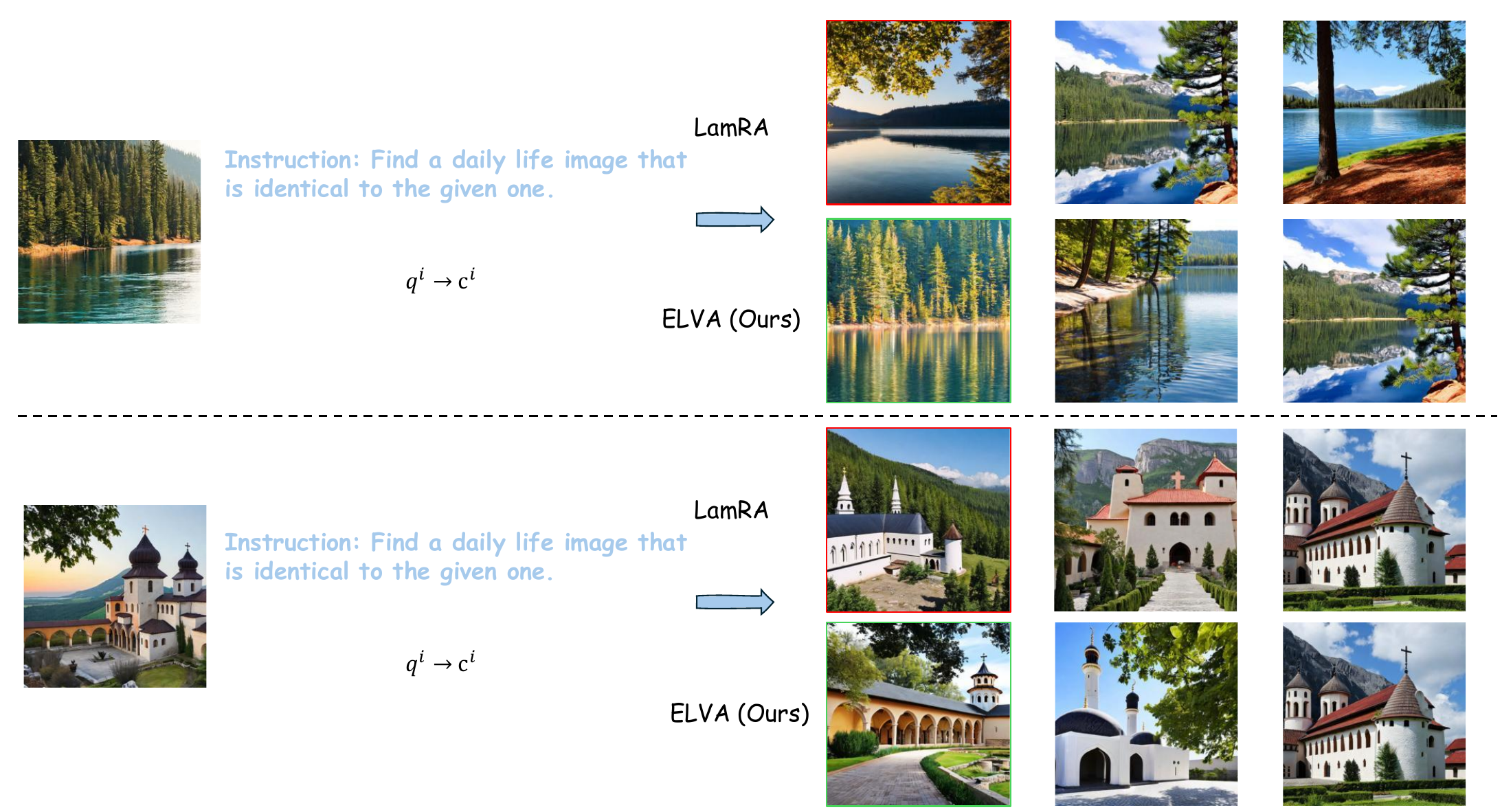} \\
  \vspace{-4pt}
  \caption{\textbf{Qualitative examples.} We show the results of our method across different retrieval tasks, with the correct result indicated by the green box. Here, $q^t$ for text queries, $q^i$ for image queries, $c^i$ for image candidates.}
 \label{fig:suppl2}
\end{figure*}

In this section, we show additional examples of successful retrieval. As illustrated in Figure \ref{fig:suppl1} and Figure \ref{fig:suppl2}, our method effectively handles a diverse range of retrieval tasks.

\begin{table*}[ht]
\centering
\caption{\textbf{Summary of the M-BEIR instructions.}
}
\vspace{-5pt}
\resizebox{.86\textwidth}{!}{  
\setlength{\tabcolsep}{0.6mm}{
  \begin{tabular}{lll}
    \toprule
    \textbf{Task} & \textbf{Dataset} & \textbf{Instruction} \\
    \midrule
    \multirow{12}{*}{$q^t \to c^i$} & \multirow{4}{*}{VisualNews} & Identify the news-related image in line with the described event. \\
     & & Display an image that best captures the following caption from the news.\\
    & & Based on the caption, provide the most fitting image for the news story. \\
    & & I want you to retrieve an image of this news caption. \\
    \cmidrule(r){2-3}
    & \multirow{4}{*}{MSCOCO} & Find me an everyday image that matches the given caption.\\
    & & Identify the image showcasing the described everyday scene. \\
    & & I want to retrieve an image of this daily life description.\\
    & & Show me an image that best captures the following common scene description.\\
     \cmidrule(r){2-3}
     & \multirow{4}{*}{Fashion200K} & Based on the following fashion description, retrieve the best matching image.\\
     & & Match the provided description to the correct fashion item photo.\\
     & & Identify the fashion image that aligns with the described product.\\
     & & You need to identify the image that corresponds to the fashion product description provided.\\
     \midrule
     \multirow{4}{*}{$q^t \to c^t $} & \multirow{4}{*}{WebQA} & Retrieve passages from Wikipedia that provide answers to the following question.\\
     & & You have to find a Wikipedia paragraph that provides the answer to the question.\\
     & & I want to find an answer to the question. Can you find some snippets that provide evidence from Wikipedia?\\
     & & I'm looking for a Wikipedia snippet that answers this question.\\
     \midrule
     \multirow{8}{*}{$q^t \to (c^i, c^t) $} & \multirow{4}{*}{EDIS} & Find a news image that matches the provided caption.\\
     & & Identify the news photo for the given caption.\\
     & & Can you pair this news caption with the right image?\\
     & & I'm looking for an image that aligns with this news caption. \\
     \cmidrule(r){2-3}
      & \multirow{4}{*}{WebQA} & Find a Wikipedia image that answers this question.\\
     & & Provide with me an image from Wikipedia to answer this question. \\
     & & I want to know the answer to this question. Please find the related Wikipedia image for me. \\
     & & You need to retrieve an evidence image from Wikipedia to address this question. \\
     \midrule
     \multirow{12}{*}{$q^i \to c^t $} & \multirow{4}{*}{VisualNews} & Find a caption for the news in the given photo.\\
     & & Based on the shown image, retrieve an appropriate news caption. \\
     & & Provide a news-related caption for the displayed image. \\
     & & I want to know the caption for this news image. \\
     \cmidrule(r){2-3}
     & \multirow{4}{*}{MSCOCO}                  & Find an image caption describing the following everyday image. \\
     & & Retrieve the caption for the displayed day-to-day image. \\
     & & Can you find a caption talking about this daily life image? \\
     & & I want to locate the caption that best describes this everyday scene image. \\
      \cmidrule(lr){2-3}
     & \multirow{4}{*}{Fashion200K} & Find a product description for the fashion item in the image.  \\
     & & Based on the displayed image, retrieve the corresponding fashion description.  \\
     & & Can you retrieve the description for the fashion item in the image?  \\
     & & I want to find a matching description for the fashion item in this image.  \\
     \midrule
     \multirow{4}{*}{$q^i \to c^i $} & \multirow{4}{*}{NIGHTS} & Find a day-to-day image that looks similar to the provided image.\\
     & & Which everyday image is the most similar to the reference image? \\
     & & Find a daily life image that is identical to the given one. \\
     & & You need to identify the common scene image that aligns most with this reference image. \\
     \midrule 
     \multirow{8}{*}{$(q^i, q^t) \to c^t $} & \multirow{4}{*}{OVEN} & Retrieve a Wikipedia paragraph that provides an answer to the given query about the image.\\
     & & Determine the Wikipedia snippet that identifies the visual entity in the image. \\
     & & I want to find a paragraph from Wikipedia that answers my question about this image. \\
     & & You have to find a Wikipedia segment that identifies this image's subject. \\
     \cmidrule(lr){2-3}
     & \multirow{4}{*}{InfoSeek} & Retrieve a Wikipedia paragraph that provides an answer to the given query about the image.\\
     & & Determine the Wikipedia snippet that matches the question of this image. \\
     & & I want to find a paragraph from Wikipedia that answers my question about this image. \\
     & & You have to find a Wikipedia segment that answers the question about the displayed image. \\
     \midrule 
     \multirow{8}{*}{$(q^i, q^t) \to c^i $} & \multirow{4}{*}{FashionIQ}  & Find a fashion image that aligns with the reference image and style note. \\
     & & With the reference image and modification instructions, find the described fashion look. \\
     & & Given the reference image and design hint, identify the matching fashion image. \\
     & & I'm looking for a similar fashion product image with the described style changes. \\
     \cmidrule(lr){2-3}
     & \multirow{4}{*}{CIRR}  & Retrieve a day-to-day image that aligns with the modification instructions of the provided image. \\
     & & Pull up a common scene image like this one, but with the modifications I asked for. \\
     & & Can you help me find a daily image that meets the modification from the given image? \\
     & & I'm looking for a similar everyday image with the described changes. \\
    \midrule
    \multirow{8}{*}{$(q^i, q^t) \to (c^i, c^t) $} & \multirow{4}{*}{OVEN} & Retrieve a Wikipedia image-description pair that provides evidence for the question of this image. \\
    & & Determine the Wikipedia image-snippet pair that clarifies the entity in this picture. \\
    & & I want to find an image and subject description from Wikipedia that answers my question about this image. \\
    & & I want to know the subject in the photo. Can you provide the relevant Wikipedia section and image? \\
    \cmidrule(lr){2-3}
    & \multirow{4}{*}{InfoSeek} & Retrieve a Wikipedia image-description pair that provides evidence for the question of this image. \\
    & & Determine the Wikipedia image-snippet pair that matches my question about this image. \\
    & & I want to find an image and subject description from Wikipedia that answers my question about this image. \\
    & & I want to address the query about this picture. Please pull up a relevant Wikipedia section and image. \\
    \bottomrule
  \end{tabular}
}
}
\vspace{-2em}
\label{tab:mbeir_instruction}
\end{table*}

\end{document}